\RequirePackage{fix-cm}
\documentclass[smallextended]{svjour3}       
\smartqed  
\usepackage{graphicx}
\begin{document}

\title{Spectrum of strange singly charmed baryons in the constituent quark model 
}


\author{Keval Gandhi{$^{1\dag}$} and Ajay Kumar Rai{$^{1\ddag}$}     
}

\authorrunning{K. Gandhi and A. K. Rai} 

\institute{\at
            $^1$Department of Applied Physics, Sardar Vallabhbhai National Institute of Technology, Surat$-$395007, Gujarat, India. \\
              $^\dag$\email{keval.physics@yahoo.com}           
                      \at
             $^\ddag$\email{raiajayk@gmail.com}           
}

\date{Received: date / Accepted: date}

\maketitle

\begin{abstract}
Excited states masses of the strange singly charmed baryons are calculated using the non-relativistic approach of hypercentral Constituent Quark Model (hCQM). The hyper-Coulomb plus screened potential is used as a confinement potential with the first order correction. The spin-spin, spin-orbit and spin-tensor interaction terms are included perturbatively. Our calculated masses are allowed to construct the Regge trajectories in both $(n_r, M^2)$  and $(J, M^2)$ planes. The mass spectra and the Regge trajectories study predict the spin-parity of $\Xi{_c(2970)^{+/0}}$, $\Xi{_c(3080)^{+/0}}$, $\Xi{_c(3123)^+}$, $\Omega{_c}(3000)^0$ and $\Omega{_c}(3119)^0$ baryons. Moreover, the strong one pion decay rates of the isodoublet states of $\Xi{_c(2645)}$, $\Xi{_c(2790)}$ and $\Xi{_c(2815)}$ are analyzed in the framework of Heavy Hadron Chiral Perturbation Theory (HHChPT). Also, the ground state magnetic moments and the radiative decay rates based on the transition magnetic moments are calculated in the framework of constituent quark model. 
\keywords{$\Xi{_c}$ and $\Omega{_c^0}$ baryons, Mass spectrum, Regge trajectories}
\end{abstract}

\section{Introduction}
\label{intro}

\begin{table}
\caption{Mass, width and $J^P$ value of the strange singly charmed baryons from PDG \cite{Tanabashi2018}.}
\label{tab:1}       
\begin{tabular}{lllllllllllllllllll}
\hline\noalign{\smallskip}
Resonance & Mass (MeV) & Width (MeV) & $J^P$ \\
\noalign{\smallskip}\hline\noalign{\smallskip}
$\Xi{_{c}^+}$ & 2467.87 $\pm$ 0.30 &$-$& ${\frac{1}{2}}^+$  \\ 
$\Xi{_{c}^0}$ & 2470.87$_{-0.31}^{+0.28}$ & $-$ & ${\frac{1}{2}}^+$ \\
$\Xi{_{c}^{\prime+}}$ & 2578.4 $\pm$ 0.5 & $-$ & ${\frac{1}{2}}^+$ \\
$\Xi{_{c}^{\prime0}}$ & 2579.2 $\pm$ 0.5 & $-$ & ${\frac{1}{2}}^+$\\
$\Xi{_c}(2645)^{+}$ & 2645.53 $\pm$ 0.31 & 2.14 $\pm$ 0.19 & ${\frac{3}{2}}^+$ \\
$\Xi{_c}(2645)^{0}$ & 2646.32 $\pm$ 0.31 & 2.35 $\pm$ 0.18 $\pm$ 0.13 & ${\frac{3}{2}}^+$ \\
$\Xi{_c}(2790)^{+}$ & 2792.0 $\pm$ 0.5 & 8.9 $\pm$ 0.6 $\pm$ 0.8 & ${\frac{1}{2}}^-$ \\
$\Xi{_c}(2790)^{0}$ & 2792.8 $\pm$ 1.2 & 10.0 $\pm$ 0.7 $\pm$ 0.8 & ${\frac{1}{2}}^-$ \\
$\Xi{_c}(2815)^{+}$ & 2816.67 $\pm$ 0.31 & 2.43 $\pm$ 0.20 $\pm$ 0.17 &${\frac{3}{2}}^-$\\
$\Xi{_c}(2815)^{0}$ & 2820.22 $\pm$ 0.32 & 2.54 $\pm$ 0.18 $\pm$ 0.17 &${\frac{3}{2}}^-$\\
$\Xi{_c}(2930)$ & 2931 $\pm$ 3 $\pm$ 5 & 36 $\pm$ 7 $\pm$ 11 & $?^?$ \\
$\Xi{_c}(2970)^{+}$ & 2969.4 $\pm$ 0.8 & 20.9$_{-3.5}^{+2.4}$ & $?^?$ \\
$\Xi{_c}(2970)^{0}$ & 2967.8 $\pm$ 0.8 & 28.1$_{-4.0}^{+3.4}$ & $?^?$ \\
$\Xi{_c}(3055)^{+}$ & 3055.9 $\pm$ 0.4 & 7.8 $\pm$ 1.2 $\pm$ 1.5 & $?^?$ \\
$\Xi{_c}(3080)^{+}$ & 3077.2 $\pm$ 0.4 & 3.6 $\pm$ 1.1 & $?^?$ \\
$\Xi{_c}(3080)^{0}$ & 3079.9 $\pm$ 1.4 & 5.6 $\pm$ 2.2 & $?^?$ \\
$\Xi{_c}(3123)^{+}$ & 3122.9 $\pm$ 1.3 $\pm$ 0.3 & 4.4 $\pm$ 3.4 $\pm$ 1.7 & $?^?$ \\
$\Omega{_{c}^0}$ & 2695.2 $\pm$ 1.7 & $-$ & ${\frac{1}{2}}^+$ \\
$\Omega{_c}(2770)^0$ & 2765.9 $\pm$ 2.0 & $-$ & ${\frac{3}{2}}^+$ \\
$\Omega{_c}(3000)^0$ & 3000.4 $\pm$ 0.2 $\pm$ 0.1 $\pm$ 0.3 & 4.5 $\pm$ 0.6 $\pm$ 0.3 & $?^?$ \\
$\Omega{_c}(3050)^0$ & 3050.2 $\pm$ 0.1 $\pm$ 0.1 $\pm$ 0.3 & $<$1.2 & $?^?$ \\
$\Omega{_c}(3065)^0$ & 3065.6 $\pm$ 0.1 $\pm$ 0.3 $\pm$ 0.3 & 3.5 $\pm$ 0.4 $\pm$ 0.2 & $?^?$ \\
$\Omega{_c}(3090)^0$ & 3090.2 $\pm$ 0.3 $\pm$ 0.5 $\pm$ 0.3 & 8.7 $\pm$ 1.0 $\pm$ 0.8 & $?^?$ \\
$\Omega{_c}(3120)^0$ & 3119.1 $\pm$ 0.3 $\pm$ 0.9 $\pm$ 0.3 & $<$2.1 & $?^?$\\
\noalign{\smallskip}\hline
\end{tabular}
\end{table}

The strange singly charmed baryons $\Xi{_c}$ and $\Omega{_c^0}$ respectively belong to the antisymmetric antitriplet and symmetric sextet SU(3) flavour representations. Recently, the Particle Data Group (PDG) listed the nine baryonic states of $\Xi{_c}$ baryon: $\Xi{_c} (2470)^{+/0}$, $\Xi{_c} (2645)^{+/0}$, $\Xi{_c} (2790)^{+/0}$, $\Xi{_c} (2815)^{+/0}$, $\Xi{_c} (2930)$, $\Xi{_c} (2970)^{+/0}$, $\Xi{_c} (3055)^{+}$, $\Xi{_c} (3080)^{+/0}$, $\Xi{_c} (3123)^{+}$; and the six of $\Omega{_c^0}$ baryon: $\Omega{_c}(2770)^0$, $\Omega{_c}(3000)^0$, $\Omega{_c}(3050)^0$, $\Omega{_c}(3065)^0$, $\Omega{_c}(3090)^0$ and $\Omega{_c}(3120)^0$ (shown in Table \ref{tab:1}) \cite{Tanabashi2018}. The masses, isospin mass splitting, spin-parity, decay widths, lifetime etc. of these baryons were measured by the ARGUS, CLEO, E687, FOCUS, $BABAR$, Belle, CDF and LHCb experimental Collaborations. The comprehensive review articles wrote a detailed informations on singly charmed baryons \cite{HXChen2017,Sonnenschein2019,Cheng20151}. Moreover, the present experimental facilities Belle II, LHCb, J-PARC and the upcoming experiment $\overline{\mbox{\sffamily P}}${\sffamily ANDA} \cite{Singh2017,Singh2016epja,Singh2016npa,Singh2015,Singh2019,Barucca2019} are expected to study singly heavy hadrons in the near future.\\

Experimentally, the ground state mass spectra of strange singly charmed baryons are now well established and many new excited states were found in the past few years. The $BABAR$ detector measured the mass and the decay width of $\Xi{_c} (2930)$ \cite{Aubert2008}. Ref. \cite{Aubert20081} observed two new resonances $\Xi{_c} (3055)^{+}$ and $\Xi{_c} (3123)^{+}$ decaying into $\Sigma_c(2455)^{++} K^-$ and $\Sigma_c(2520)^{++} K^-$ final states respectively. These were also seen by the Belle Collaboration \cite{Kato2014}. Two isodoublet baryonic states of $\Xi{_c} (2970)$ and $\Xi{_c} (3080)$ have been measured by BABAR \cite{Aubert20081} and the Belle \cite{Lesiak2008,Kato2016} Collaborations. In 2017, the LHCb Collaboration \cite{Aaij2017} observed the five new narrow resonances $\Omega{_c}(3000)^0$, $\Omega{_c}(3050)^0$, $\Omega{_c}(3066)^0$, $\Omega{_c}(3090)^0$ and $\Omega{_c}(3119)^0$ decaying into $\Xi{_{c}^+} K^-$. The spin-parity of these excited strange charmed baryons are not confirmed yet, as shown in Table \ref{tab:1}. The $J^P$ value assignment of the heavy-light hadrons is very crucial, because can help to probe their properties as: form factors, decay widths, branching fractions, hyperfine splittings etc.. That allows to look back to the theory and the phenomenological study. In Ref. \cite{BChen2017} the authors solved light quark cluster-heavy quark picture with the help of the Cornell potential in the framework of non-relativistic constitute quark model. They calculate the masses of radial excited states 2S and 3S as well the orbital excited states 1P, 2P and 1D of charmed ($\Lambda_c^+, \Sigma_c$) and charmed-strange ($\Xi_c, \Xi_c^{\prime}$) baryons by considering an equal up $(u)$ and down $(d)$ quarks masses. So they can$'$t distinguished the mass spectra of an isodoublet $\Xi_c^0$ and $\Xi_c^+$ (or $\Xi_c^{\prime0}$ and $\Xi_c^{\prime+}$) baryons. Z. Shah \textit{et al.} \cite{Shah2016epja,Shah2016cpc} used non-relativistic approach of hypercentral Constituent Quark Model (hCQM) and calculate the excited state masses of singly charmed baryons. The hyper-Coulomb potential with linear potential of power index $\nu$, varying from 0.5 to 2.0, is used as a long range spin-independent potential. The mass spectra of isodoublet $\Xi_c$ baryons are calculated separately by considering an unequal $u$ and $d$ quarks masses. Ref. \cite{Yoshida2015} solved the three-body bound state problems using Gaussian expansion method in the non-relativistic framework of a constituent quark model. The harmonic oscillator potential is used as a scalar confinement potential. Choosing an equal $u$ and $d$ quarks masses the excited state mass spectra of $\Lambda_c^+$, $\Sigma_c$ and $\Omega_c^0$ baryons are calculated in the two different excited modes, say $\lambda$ and $\rho$. The non-relativistic Hamiltonian quark model inspired by Isgur and Karl model calculates the masses of baryons containing one, two and three heavy flavor quarks in Ref. \cite{Roberts2008} with equal $u$ and $d$ quarks masses. Additionally, they introduced the Heavy Quark Effective Theory (HQET) that predicts the degeneracy in baryonic multiplets containing a single heavy quark in terms of quark model quantum numbers. D. Ebert \textit{et al.} \cite{Ebert2011,Ebert2008} calculated the mass spectra of heavy baryons in the Quantum Chromodynamics (QCD) motivated relativistic heavy-quark-light-diquark model. Ref. \cite{BChen2015} derived a mass formula analytically for the excited heavy-light hadrons within the framework of relativistic flux tube model, where the heavy quark-light diquark picture is considered. Such formula is applied to determine the masses of $\Lambda_Q$ and $\Xi_Q$ (Q = $c$/$b$) baryons. The authors of Ref. \cite{Yamaguchi2015} obtained the masses of one heavy quark baryons in the heavy quark limit. H.-X. Chen \textit{et al.} \cite{Chen2015} studied the mass spectra of $P-$wave charmed baryons using the QCD sum rule in the framework of HQET. Ref. \cite{Valcarce2008} used Faddeev method  and solved the three-quark problem in momentum space and calculated the masses of singly, doubly and triply heavy baryons. Ref. \cite{Guo2008} derived some useful mass relations for hadrons using quasilinear Regge trajectory ansatz and make a $J^P$ assignments of $\Xi_c(2980)$, $\Xi_c(3055)$, $\Xi_c(3077)$ and $\Xi_c(3123)$ baryons. Ref. \cite{Rubio2015} determined the ground state and the first excited state masses of singly and doubly charmed baryons of spin 1/2 and 3/2 with positive and negative parity. N. Mathur \textit{et al.} \cite{Mathur2002} used quenched lattice non-relativistic QCD approach and computed the mass spectrum of charmed and bottom baryons.\\

The strange singly charmed baryons contain one charm quark $(c)$, one strange quark $(s)$ and one light quark $(u, d$ and $ s)$ out of these three quarks. This provides an excellent ground for testing the heavy quark symmetry of the heavy quark and the chiral symmetry of the light quarks in a low energy regime. The strong decay of singly charmed baryons were conveniently studied in the framework of Heavy Hadron Chiral Perturbation Theory (HHChPT) by which the heavy quark symmetry and the chiral symmetry are incorporated. The chiral Lagrangians corresponding to the strong decay of singly charmed baryons were constructed in the Refs. \cite{Pirjol1997,Cheng2007,Cheng2015} for the $S$, $P$ and $D$-wave transitions. The electromagnetic or weak decays of singly charmed baryons are not well established and their evidences are not found experimentally yet. Several approaches are used for computing the properties of singly charmed baryons: heavy baryon chiral perturbation theory (HBChPT) \cite{Wang2019,Meng2018,Jiang2015}, $SU(3)$ chiral quark-soliton model \cite{JYKim2019}, pion mean-field approach \cite{HCKim2019}, chiral structure model \cite{Kawakami2018}, covariant chiral perturbation theory \cite{Shi2018}, large $N_c$ limit \cite{Yang2018}, chiral quark model \cite{Wang20171}, non-relativistic constituent quark model \cite{Gandhi2018,Majethiya2009,Patel2008,Albertus2005}, lattice QCD \cite{Can2015,Can2014}, the bag model \cite{Bernotas2013}, light-cone QCD sum rule \cite{Aliev2012,Aliev2009}, $SU(4)$ chiral constituent quark model ($\chi$CQM) \cite{Sharma2010}, relativistic constituent quark model \cite{Faessler2006,Ivanov1999}, light front quark model \cite{Tawfiq1998}, the $^3P_0$ model \cite{Chen2007} and others.\\ 

Our effort is to calculate the radial ($2S$-$6S$) and the orbital ($1P$-$5P$, $1D$-$5D$, $1F$-$4F$, $1G$-$3G$ and $1H$-$2H$) excited states masses of strange singly charmed baryons, for orbital quantum number $L$ = 0, 1, 2, 3, 4, 5, in the framework of non-relativistic hypercentral Constituent Quark Model (hCQM). The vector Coulomb potential with an additional screening component is used as a scalar confinement potential with the first order correction to give relativistic effect of order $\cal{O}$$(1/m)$. Such calculated masses are allowed to construct the Regge trajectories in $(n_r, M^2)$  and $(J, M^2)$ planes.\\

This paper is organized as follows. The hCQM and the potential model is described in brief in section \ref{sec:2}. We discuss the mass spectra and construct the Regge trajectories in section \ref{sec:3}. The strong one pion decay rates of the isodoublet $\Xi{_c(2645)}$, $\Xi{_c(2790)}$ and $\Xi{_c(2815)}$ baryons are determined separately by using their masses obtained in the present study and PDG-2018, and the properties like magnetic moments and the radiative decay widths of the ground state strange charmed baryons are presented in section \ref{sec:4}. In the last section \ref{sec:5} we summarize our present work.

\begin{table*}
\caption{Predicted masses of the radial excited states of strange singly charmed baryons (in GeV).}
\label{tab:2}       
\scalebox{1.0}{
\begin{tabular}{llllllllllllllllllllllll}
\hline\noalign{\smallskip}
Particle & State & $J^P$ & Present & \cite{Shah2016epja} & \cite{Ebert2011} & \cite{BChen2017} & \cite{BChen2015} & \cite{Ebert2008} & \cite{Yamaguchi2015} & \cite{Valcarce2008} & \cite{Yoshida2015} & \cite{Roberts2008} & PDG \cite{Tanabashi2018}\\
\noalign{\smallskip}\hline\noalign{\smallskip}
$\Xi{_c^0}$ & $1S$ & ${\frac{1}{2}}^+$ & 2.471 & 2.470 & 2.476 & 2.470 & 2.467 & 2.481 & 2.466 & 2.471 & & 2.492 & 2.47091 $\pm$ 0.00025\\
& $2S$ & & 2.964 & 2.940 & 2.959 & 2.940 & 2.959 & 2.923 & 2.924 & 3.137 &&& 2.9707 $\pm$ 0.0022 \\
& $3S$ & & 3.358 & 3.339 & 3.323 & 3.265 & 3.325 & 3.313 & 3.183 &&& \\
& $4S$ & & 3.720 & 3.726 & 3.632 & & 3.629\\
& $5S$ & & 4.064 & 4.107 & 3.909\\
& $6S$ & & 4.394 &  & 4.166\\
\cline{2-14}
&$1S$ & ${\frac{3}{2}}^+$ & 2.648 & 2.610 & 2.654 &&&&& 2.642 && 2.650 & 2.64838 $\pm$ 0.00021  \\
& $2S$ & & 3.080 & 3.038 &&&&&& 3.071 && 2.984 & 3.0799 $\pm$ 0.0014  \\
& $3S$ & & 3.424 & 3.399 \\
& $4S$ & & 3.763 & 3.766 \\
& $5S$ & & 4.093 & 4.136 \\
& $6S$ & & 4.415 & \\
\noalign{\smallskip}\hline
$\Xi{_c^+}$ & $1S$ & ${\frac{1}{2}}^+$ & 2.467 & 2.467 &&&&&&&&& 2.46793 $\pm$ 0.00018  \\
& $2S$ & & 2.969 & 2.956 &&&&&&&&& 2.968 $\pm$ 0.0026 \\
& $3S$ & & 3.369 & 3.370  \\
& $4S$ & & 3.738 & 3.772 \\
& $5S$ & & 4.088 & 4.167  \\
& $6S$ & & 4.424 \\
\cline{2-14}
& $1S$ & ${\frac{3}{2}}^+$ & 2.646 & 2.619 &&&&&&&&& 2.64557 $\pm$ 0.00026\\
& $2S$ & & 3.086 & 3.061 &&&&&&&&& 3.0772 $\pm$ 0.0004 \\
& $3S$ & & 3.436 & 3.435 \\
& $4S$ & & 3.781 & 3.815 \\
& $5S$ & & 4.117 & 4.198 \\
& $6S$ & & 4.445 \\
\noalign{\smallskip}\hline
$\Omega{_c^0}$ & $1S$ & ${\frac{1}{2}}^+$ & 2.696 & 2.695 & 2.698 &&& 2.698 & 2.718 & 2.699 & 2.731 & 2.718 & 2.6952 $\pm$ 0.0017 \\
& $2S$ & & 3.165 & 3.164 & 3.088 &&&3.065& 3.152 & 3.159 & 3.227 & 3.152 & 3.1191 $\pm$ 0.0003\\
& $3S$ & & 3.540 & 3.561 & 3.489 &&&& 3.275 & & 3.292 \\
& $4S$ & & 3.895 & 3.953 & 3.814 &&&&  3.299 & \\
& $5S$ & & 4.238 & 4.343 & 4.102 \\
& $6S$ & & 4.569 &  \\
\cline{2-14}
& $1S$ & ${\frac{3}{2}}^+$ & 2.766 & 2.745 & 2.768 &&& 2.768 & 2.766 & 2.767 & 2.779 & 2.776 & 2.7659 $\pm$ 0.002 \\
& $2S$ & & 3.208 & 3.197 & 3.123 &&& 3.119 & 3.190 & 3.202 & 3.257 & 3.190 \\
& $3S$ & & 3.564 & 3.580 & 3.510 &&&& 3.280 && 3.285 \\
& $4S$ & & 3.911 & 3.966 & 3.830 &&&& 3.321  \\
& $5S$ & & 4.248 & 4.352 & 4.114 \\
& $6S$ & & 4.577 & \\
\noalign{\smallskip}\hline
\end{tabular}}
\end{table*}

\begin{table}
\caption{Predicted masses of the orbital excited states of $\Xi{_{c}^0}$ baryon ($P$ and $D$ wave) (in GeV).}
\label{tab:3}       
\begin{tabular}{lllllllllllllllllll}
\hline\noalign{\smallskip}
State & $J^P$ & Present & \cite{Shah2016epja} & \cite{Ebert2011} & \cite{BChen2017} & \cite{BChen2015} & \cite{Ebert2008} & \cite{Yamaguchi2015} &  \cite{Valcarce2008} & \cite{Chen2015} & \cite{Roberts2008} & PDG \cite{Tanabashi2018} \\
\noalign{\smallskip}\hline\noalign{\smallskip}
$1P$ & ${\frac{1}{2}}^-$ & 2.828 & 2.796 & 2.792 & 2.793 & 2.779 & 2.801 & 2.773 & 2.799 & 2.790 & 2.763 & 2.7941 $\pm$ 0.0005 \\
& ${\frac{3}{2}}^-$ & 2.820 & 2.781 & 2.819 & 2.820 & 2.814 & 2.820 & 2.783 & & 2.830 & 2.784 & 2.82026 $\pm$ 0.00027  \\
& ${\frac{1}{2}}^-$ & 2.832 & 2.803 &&&&&& 2.902 & & 2.859\\
& ${\frac{3}{2}}^-$ & 2.824 & 2.788 &&&&&&&& 2.871 \\
& ${\frac{5}{2}}^-$ & 2.814 & 2.769 &&&&&&&& 2.905 \\
\noalign{\smallskip}\hline
$2P$ & ${\frac{1}{2}}^-$ & 3.191 & 3.170 & 3.179 & 3.140 & 3.195 & 3.186 && 3.004 & &\\
& ${\frac{3}{2}}^-$ & 3.184 & 3.154 & 3.201 & 3.164 & 3.204 & 3.199\\
& ${\frac{1}{2}}^-$ & 3.195 & 3.178 & \\
& ${\frac{3}{2}}^-$ & 3.188 & 3.162 & \\
& ${\frac{5}{2}}^-$ & 3.177 & 3.140 &&&&&&&& 2.985 \\
\noalign{\smallskip}\hline
$3P$ & ${\frac{1}{2}}^-$ & 3.541 & 3.546 & 3.500 & & 3.521 \\
& ${\frac{3}{2}}^-$ & 3.533 & 3.527 & 3.519 & & 3.525 \\
& ${\frac{1}{2}}^-$ & 3.545 & 3.555 &  \\
& ${\frac{3}{2}}^-$ & 3.537 & 3.536 & \\
& ${\frac{5}{2}}^-$ & 3.527 & 3.511 & \\
\noalign{\smallskip}\hline
$4P$ & ${\frac{1}{2}}^-$ & 3.879 & 3.919 & 3.785 \\
& ${\frac{3}{2}}^-$ & 3.871 & 3.899 & 3.804 \\
& ${\frac{1}{2}}^-$ & 3.883 & 3.929 \\
& ${\frac{3}{2}}^-$ & 3.875 & 3.909 \\
& ${\frac{5}{2}}^-$ & 3.865 & 3.882 \\
\noalign{\smallskip}\hline
$5P$ & ${\frac{1}{2}}^-$ & 4.206 & 4.293 & 4.048 \\
& ${\frac{3}{2}}^-$ & 4.199 & 4.272 & 4.066  \\
& ${\frac{1}{2}}^-$ & 4.210 & 4.304 \\
& ${\frac{3}{2}}^-$ & 4.203 & 4.283 \\
& ${\frac{5}{2}}^-$ & 4.193 & 4.254 \\
\noalign{\smallskip}\hline
$1D$ & ${\frac{3}{2}}^+$ & 3.116 & 3.080 & 3.059 & 3.033 & 3.055 & 3.030 & 3.012 & & & & \\
& ${\frac{5}{2}}^+$ & 3.103 & 3.054 & 3.076 & 3.040 & 3.076 & 3.042 & 3.004 & 3.049 && 2.995 & \\
& ${\frac{1}{2}}^+$ & 3.131 & 3.108 &&&&&&&&&  \\
& ${\frac{3}{2}}^+$ & 3.121 & 3.089 & \\
& ${\frac{5}{2}}^+$ & 3.108 & 3.064 &&&&&& 3.132 \\
& ${\frac{7}{2}}^+$ & 3.092 & 3.032 &&&&&&& 3.100 \\
\noalign{\smallskip}\hline
$2D$ & ${\frac{3}{2}}^+$ & 3.464 & 3.450 & 3.388 & & 3.407 &3.411 \\
& ${\frac{5}{2}}^+$ & 3.452 & 3.422 & 3.407 & & 3.416 & 3.413\\
& ${\frac{1}{2}}^+$ & 3.478 & 3.480 \\
& ${\frac{3}{2}}^+$ & 3.469 & 3.460 \\
& ${\frac{5}{2}}^+$ & 3.457 & 3.432 \\
& ${\frac{7}{2}}^+$ & 3.442 & 3.398 \\
\noalign{\smallskip}\hline
$3D$ & ${\frac{3}{2}}^+$ & 3.804 &  & 3.678 \\
& ${\frac{5}{2}}^+$ & 3.792 &  & 3.699 \\
& ${\frac{1}{2}}^+$ & 3.817 & & \\
& ${\frac{3}{2}}^+$ & 3.308 & & \\
& ${\frac{5}{2}}^+$ & 3.796 & & \\
& ${\frac{7}{2}}^+$ & 3.782 & & \\
\noalign{\smallskip}\hline
$4D$ & ${\frac{3}{2}}^+$ & 4.132 &  & 3.945 \\
& ${\frac{5}{2}}^+$ & 4.121 &  & 3.965 \\
& ${\frac{1}{2}}^+$ & 4.144 & & \\
& ${\frac{3}{2}}^+$ & 4.136 & & \\
& ${\frac{5}{2}}^+$ & 4.125 & & \\
& ${\frac{7}{2}}^+$ & 4.112 & & \\
\noalign{\smallskip}\hline
$5D$ & ${\frac{3}{2}}^+$ & 4.450 \\
& ${\frac{5}{2}}^+$ & 4.442 \\
& ${\frac{1}{2}}^+$ & 4.459 \\
& ${\frac{3}{2}}^+$ & 4.453 \\
& ${\frac{5}{2}}^+$ & 4.445 \\
& ${\frac{7}{2}}^+$ & 4.435 \\
\noalign{\smallskip}\hline
\end{tabular}
\end{table}

\begin{table}
\addtocounter{table}{-1}
\caption{Predicted masses of the orbital excited states of $\Xi{_{c}^0}$ baryon ($F$, $G$ and $H$ wave) (in GeV).}
\label{tab:3}       
\begin{tabular}{lllllllllllllllllll}
\hline\noalign{\smallskip}
State & $J^P$ & Present & \cite{Shah2016epja} & \cite{Ebert2011} & \cite{BChen2015} & \cite{Ebert2008}\\
\noalign{\smallskip}\hline\noalign{\smallskip}
$1F$ & ${\frac{5}{2}}^-$ & 3.388 & 3.352 & 3.278 & 3.286 & 3.219\\
& ${\frac{7}{2}}^-$ & 3.369 & 3.316 & 3.292 & 3.302 & 3.208 \\
& ${\frac{3}{2}}^-$ & 3.408 & 3.392 \\
& ${\frac{5}{2}}^-$ & 3.393 & 3.363 \\
& ${\frac{7}{2}}^-$ & 3.375 & 3.327 \\
& ${\frac{9}{2}}^-$ & 3.358 & 3.283 \\
\noalign{\smallskip}\hline
$2F$ & ${\frac{5}{2}}^-$ & 3.727 &  & 3.575 \\
& ${\frac{7}{2}}^-$ & 3.710 &  & 3.592 \\
& ${\frac{3}{2}}^-$ & 3.745 & & \\
& ${\frac{5}{2}}^-$ & 3.732 & &  \\
& ${\frac{7}{2}}^-$ & 3.715 & & \\
& ${\frac{9}{2}}^-$ & 3.695 & & \\
\noalign{\smallskip}\hline
$3F$ & ${\frac{5}{2}}^-$ & 4.055 &  & 3.845 \\
& ${\frac{7}{2}}^-$ & 4.042 &  & 3.865 \\
& ${\frac{3}{2}}^-$ & 4.069 & & \\
& ${\frac{5}{2}}^-$ & 4.059 & & \\
& ${\frac{7}{2}}^-$ & 4.046 & & \\
& ${\frac{9}{2}}^-$ & 4.030 & & \\
\noalign{\smallskip}\hline
$4F$ & ${\frac{5}{2}}^-$ & 4.376 &  & 4.098 \\
& ${\frac{7}{2}}^-$ & 4.364 &  & 4.120 \\
& ${\frac{3}{2}}^-$ & 4.388 & &  \\
& ${\frac{5}{2}}^-$ & 4.379 & & \\
& ${\frac{7}{2}}^-$ & 4.368 & & \\
& ${\frac{9}{2}}^-$ & 4.354 & & \\
\noalign{\smallskip}\hline
$1G$ & ${\frac{7}{2}}^+$ & 3.647 &  & 3.469 \\
& ${\frac{9}{2}}^+$ & 3.627 &  & 3.483 \\
& ${\frac{5}{2}}^+$ & 3.670 & & \\
& ${\frac{7}{2}}^+$ & 3.653 & & \\
& ${\frac{9}{2}}^+$ & 3.632 & & \\
& ${\frac{11}{2}}^+$ & 3.608 & &  \\
\noalign{\smallskip}\hline
$2G$ & ${\frac{7}{2}}^+$ & 3.981 &  & 3.745 \\
& ${\frac{9}{2}}^+$ & 3.960 &  & 3.763 \\
& ${\frac{5}{2}}^+$ & 4.004 & & \\
& ${\frac{7}{2}}^+$ & 3.986 & & \\
& ${\frac{9}{2}}^+$ & 3.965 & & \\
& ${\frac{11}{2}}^+$ & 3.940 & & \\
\noalign{\smallskip}\hline
$3G$ & ${\frac{7}{2}}^+$ & 4.303 \\
& ${\frac{9}{2}}^+$ & 4.285 \\
& ${\frac{5}{2}}^+$ & 4.323 \\
& ${\frac{7}{2}}^+$ & 4.308 \\
& ${\frac{9}{2}}^+$ & 4.290 \\
& ${\frac{11}{2}}^+$ & 4.268 \\
\noalign{\smallskip}\hline
$1H$ & ${\frac{9}{2}}^-$ & 3.905 &  & 3.643 \\
& ${\frac{11}{2}}^-$ & 3.877 &  & 3.658 \\
& ${\frac{7}{2}}^-$ & 3.936 & & \\
& ${\frac{9}{2}}^-$ & 3.911 & & \\
& ${\frac{11}{2}}^-$ & 3.883 & & \\
& ${\frac{13}{2}}^-$ & 3.850 & & \\
\noalign{\smallskip}\hline
$2H$ & ${\frac{9}{2}}^-$ & 4.229 \\
& ${\frac{11}{2}}^-$ & 4.204 \\
& ${\frac{7}{2}}^-$ & 4.256 \\
& ${\frac{9}{2}}^-$ & 4.234 \\
& ${\frac{11}{2}}^-$ & 4.210 \\
& ${\frac{13}{2}}^-$ & 4.181 \\
\noalign{\smallskip}\hline
\end{tabular}
\end{table}

\begin{table}
\caption{Predicted masses of the orbital excited states of $\Xi{_{c}^+}$ baryon ($P$ and $D$ wave) (in GeV).}
\label{tab:4}       
\begin{tabular}{llllllll}
\hline\noalign{\smallskip}
State & $J^P$ & Present & \cite{Shah2016epja} & PDG \cite{Tanabashi2018} \\
\noalign{\smallskip}\hline\noalign{\smallskip}
$1P$ & ${\frac{1}{2}}^-$ & 2.830 & 2.810 & 2.7924 $\pm$ 0.0005 \\
& ${\frac{3}{2}}^-$ & 2.822 & 2.794 & 2.81673 $\pm$ 0.00021 \\
& ${\frac{1}{2}}^-$ & 2.834 & 2.818 \\
& ${\frac{3}{2}}^-$ & 2.826 & 2.802\\
& ${\frac{5}{2}}^-$ & 2.815 & 2.780 \\
\noalign{\smallskip}\hline
$2P$ & ${\frac{1}{2}}^-$ & 3.200 & 3.198 \\
& ${\frac{3}{2}}^-$ & 3.192 & 3.180 \\
& ${\frac{1}{2}}^-$ & 3.204 & 3.208 \\
& ${\frac{3}{2}}^-$ & 3.196 & 3.189 \\
& ${\frac{5}{2}}^-$ &  3.185 & 3.165 \\
\noalign{\smallskip}\hline
$3P$ & ${\frac{1}{2}}^-$ & 3.556 & 3.586 \\
& ${\frac{3}{2}}^-$ & 3.547 & 3.566 \\
& ${\frac{1}{2}}^-$ & 3.560 & 3.596 \\
& ${\frac{3}{2}}^-$ & 3.551 & 3.576 \\
& ${\frac{5}{2}}^-$ & 3.540 & 3.549  \\
\noalign{\smallskip}\hline
$4P$ & ${\frac{1}{2}}^-$ & 3.899 & 3.974 \\
& ${\frac{3}{2}}^-$ & 3.891 & 3.952 \\
& ${\frac{1}{2}}^-$ & 3.903 & 3.985 \\
& ${\frac{3}{2}}^-$ & 3.895 & 3.963 \\
& ${\frac{5}{2}}^-$ & 3.885 & 3.933 \\
\noalign{\smallskip}\hline
$5P$ & ${\frac{1}{2}}^-$ & 4.233 & 4.360 \\
& ${\frac{3}{2}}^-$ & 4.225 & 4.338 \\
& ${\frac{1}{2}}^-$ & 4.237 & 4.372 \\
& ${\frac{3}{2}}^-$ & 4.229 & 4.349 \\
& ${\frac{5}{2}}^-$ & 4.219 & 4.319 \\
\noalign{\smallskip}\hline
$1D$ & ${\frac{3}{2}}^+$ & 3.123 & 3.104 & 3.1229 $\pm$ 0.0013 \\
& ${\frac{5}{2}}^+$ & 3.109 & 3.077 \\
& ${\frac{1}{2}}^+$ & 3.138 & 3.134 \\
& ${\frac{3}{2}}^+$ & 3.128 & 3.114 \\
& ${\frac{5}{2}}^+$ & 3.114 & 3.087 \\
& ${\frac{7}{2}}^+$ & 3.097 & 3.053 \\
\noalign{\smallskip}\hline
$2D$ & ${\frac{3}{2}}^+$ & 3.478 & 3.485 \\
& ${\frac{5}{2}}^+$ & 3.465 & 3.458 \\
& ${\frac{1}{2}}^+$ & 3.493 & 3.516 \\
& ${\frac{3}{2}}^+$ & 3.483 & 3.495 \\
& ${\frac{5}{2}}^+$ & 3.470 & 3.468 \\
& ${\frac{7}{2}}^+$ & 3.453 & 3.434 \\
\noalign{\smallskip}\hline
$3D$ & ${\frac{3}{2}}^+$ & 3.823 \\
& ${\frac{5}{2}}^+$ & 3.810 \\
& ${\frac{1}{2}}^+$ & 3.836 \\
& ${\frac{3}{2}}^+$ & 3.827 \\
& ${\frac{5}{2}}^+$ & 3.815 \\
& ${\frac{7}{2}}^+$ & 3.800 \\
\noalign{\smallskip}\hline
$4D$ & ${\frac{3}{2}}^+$ & 4.157  \\
& ${\frac{5}{2}}^+$ & 4.146 \\
& ${\frac{1}{2}}^+$ & 4.170 \\
& ${\frac{3}{2}}^+$ & 4.162 \\
& ${\frac{5}{2}}^+$ & 4.150 \\
& ${\frac{7}{2}}^+$ & 4.136 \\
\noalign{\smallskip}\hline
$5D$ & ${\frac{3}{2}}^+$ & 4.483 \\
& ${\frac{5}{2}}^+$ & 4.472 \\
& ${\frac{1}{2}}^+$ & 4.494 \\
& ${\frac{3}{2}}^+$ & 4.487 \\
& ${\frac{5}{2}}^+$ & 4.476  \\
& ${\frac{7}{2}}^+$ & 4.463 \\
\noalign{\smallskip}\hline
\end{tabular}
\end{table}

\begin{table}
\addtocounter{table}{-1}
\caption{Predicted masses of the orbital excited states of $\Xi{_{c}^+}$ baryon ($F$, $G$ and $H$ wave) (in GeV).}
\label{tab:4}       
\begin{tabular}{llllllll}
\hline\noalign{\smallskip}
State & $J^P$ & Present & \cite{Shah2016epja} \\
\noalign{\smallskip}\hline\noalign{\smallskip}
$1F$ & ${\frac{5}{2}}^-$ & 3.380 & 3.394 \\
& ${\frac{7}{2}}^-$ & 3.357 & 3.354 \\
& ${\frac{3}{2}}^-$ & 3.420 & 3.432 \\
& ${\frac{5}{2}}^-$ & 3.405 & 3.406 \\
& ${\frac{7}{2}}^-$ & 3.386 & 3.366 \\
& ${\frac{9}{2}}^-$ & 3.363 & 3.318 \\
\noalign{\smallskip}\hline
$2F$ & ${\frac{5}{2}}^-$ & 3.744 \\
& ${\frac{7}{2}}^-$ & 3.727 \\
& ${\frac{3}{2}}^-$ & 3.762 \\
& ${\frac{5}{2}}^-$ & 3.749 \\
& ${\frac{7}{2}}^-$ & 3.732 \\
& ${\frac{9}{2}}^-$ & 3.712 \\
\noalign{\smallskip}\hline
$3F$ & ${\frac{5}{2}}^-$ & 4.079  \\
& ${\frac{7}{2}}^-$ & 4.065 \\
& ${\frac{3}{2}}^-$ & 4.094 \\
& ${\frac{5}{2}}^-$ & 4.083 \\
& ${\frac{7}{2}}^-$ & 4.069  \\
& ${\frac{9}{2}}^-$ & 4.052 \\
\noalign{\smallskip}\hline
$4F$ & ${\frac{5}{2}}^-$ & 4.407  \\
& ${\frac{7}{2}}^-$ & 4.393 \\
& ${\frac{3}{2}}^-$ & 4.422 \\
& ${\frac{5}{2}}^-$ & 4.411 \\
& ${\frac{7}{2}}^-$ & 4.397 \\
& ${\frac{9}{2}}^-$ & 4.381 \\
\noalign{\smallskip}\hline
$1G$ & ${\frac{7}{2}}^+$ & 3.667 \\
& ${\frac{9}{2}}^+$ & 3.642 \\
& ${\frac{5}{2}}^+$ & 3.694 \\
& ${\frac{7}{2}}^+$ & 3.673 \\
& ${\frac{9}{2}}^+$ & 3.648  \\
& ${\frac{11}{2}}^+$ & 3.619 \\
\noalign{\smallskip}\hline
$2G$ & ${\frac{7}{2}}^+$ & 4.002 \\
& ${\frac{9}{2}}^+$ & 3.981 \\
& ${\frac{5}{2}}^+$ & 4.024 \\
& ${\frac{7}{2}}^+$ & 4.007 \\
& ${\frac{9}{2}}^+$ & 3.987 \\
& ${\frac{11}{2}}^+$ & 3.963  \\
\noalign{\smallskip}\hline
$3G$ & ${\frac{7}{2}}^+$ & 4.332 \\
& ${\frac{9}{2}}^+$ & 4.312 \\
& ${\frac{5}{2}}^+$ & 4.353 \\
& ${\frac{7}{2}}^+$ & 4.337 \\
& ${\frac{9}{2}}^+$ & 4.317 \\
& ${\frac{11}{2}}^+$ & 4.294 \\
\noalign{\smallskip}\hline
$1H$ & ${\frac{9}{2}}^-$ & 3.926 \\
& ${\frac{11}{2}}^-$ & 3.896 \\
& ${\frac{7}{2}}^-$ & 3.958 \\
& ${\frac{9}{2}}^-$ & 3.932 \\
& ${\frac{11}{2}}^-$ & 3.903 \\
& ${\frac{13}{2}}^-$ & 3.869 \\
\noalign{\smallskip}\hline
$2H$ & ${\frac{9}{2}}^-$ & 4.255 \\
& ${\frac{11}{2}}^-$ & 4.230 \\
& ${\frac{7}{2}}^-$ & 4.281 \\
& ${\frac{9}{2}}^-$ & 4.260 \\
& ${\frac{11}{2}}^-$ & 4.236 \\
& ${\frac{13}{2}}^-$ & 4.208 \\
\noalign{\smallskip}\hline
\end{tabular}
\end{table}

\begin{table}
\caption{Predicted masses of the orbital excited states of $\Omega{_{c}^0}$ baryon ($P$ and $D$ wave) (in GeV).}
\label{tab:5}       
\begin{tabular}{lllllllllllllll}
\hline\noalign{\smallskip}
State & $J^P$ & Present & \cite{Shah2016epja} & \cite{Ebert2011} & \cite{Ebert2008} & \cite{Yamaguchi2015} & \cite{Valcarce2008} & \cite{Yoshida2015} & \cite{Chen2015} & \cite{Roberts2008} & PDG \cite{Tanabashi2018} \\
\noalign{\smallskip}\hline\noalign{\smallskip}
$1P$ & ${\frac{1}{2}}^-$ & 2.976 & 3.041 & 3.055  & 3.025 & 3.046 & 3.035 & 3.030 & 3.250 & 2.977 & 3.0004 $\pm$ 0.00021   \\
& ${\frac{3}{2}}^-$ & 2.968 & 3.024 & 3.054 & 3.026 & 3.056 & 3.125 & 3.033 & 3.260 & 2.986 \\
& ${\frac{1}{2}}^-$ & 2.979 & 3.050 & 2.966 & 3.020 & & & 3.048 & & 2.970 \\
& ${\frac{3}{2}}^-$ & 2.972 & 3.033 & 3.029 & 2.998 & 2.986 && 3.056 & 3.270 & 2.994\\
& ${\frac{5}{2}}^-$ & 2.962 & 3.010 & 3.051 & 3.022 & 3.014 && 3.057 & 3.320 & 3.014\\
\noalign{\smallskip}\hline
$2P$ & ${\frac{1}{2}}^-$ & 3.341 & 3.427 & 3.435 & 3.376 & & \\
& ${\frac{3}{2}}^-$ & 3.334 & 3.408 & 3.433 & 3.374 && \\
& ${\frac{1}{2}}^-$ & 3.345 & 3.436 & 3.384 & 3.371\\
& ${\frac{3}{2}}^-$ & 3.337 & 3.417 & 3.415 & 3.350\\
& ${\frac{5}{2}}^-$ & 3.328 & 3.393 & 3.427 & 3.365 && \\
\noalign{\smallskip}\hline
$3P$ & ${\frac{1}{2}}^-$ & 3.694 & 3.813 & 3.754 \\
& ${\frac{3}{2}}^-$ & 3.687 & 3.793 & 3.752  \\
& ${\frac{1}{2}}^-$ & 3.699 & 3.823 & 3.717 \\
& ${\frac{3}{2}}^-$ & 3.690 & 3.803 & 3.737  \\
& ${\frac{5}{2}}^-$ & 3.681 & 3.777 & 3.744 &&&\\
\noalign{\smallskip}\hline
$4P$ & ${\frac{1}{2}}^-$ & 4.036 & 4.199 & 4.037 \\
& ${\frac{3}{2}}^-$ & 4.029 & 4.179 & 4.036 \\
& ${\frac{1}{2}}^-$ & 4.040 & 4.209 & 4.009 \\
& ${\frac{3}{2}}^-$ & 4.033 & 4.189 & 4.023 \\
& ${\frac{5}{2}}^-$ & 4.023 & 4.162 & 4.028 \\
\noalign{\smallskip}\hline
$5P$ & ${\frac{1}{2}}^-$ & 4.369 & 4.587 \\
& ${\frac{3}{2}}^-$ & 4.362 & 4.565 \\
& ${\frac{1}{2}}^-$ & 4.373 & 4.598 \\
& ${\frac{3}{2}}^-$ & 4.366 & 4.576 \\
& ${\frac{5}{2}}^-$ & 4.356 & 4.547 \\
\noalign{\smallskip}\hline
$1D$ & ${\frac{3}{2}}^+$ & 3.264 & 3.325 & 3.298 & 3.215 & 3.262\\
& ${\frac{5}{2}}^+$ & 3.251 & 3.299 & 3.297 & 3.218 & 3.273& &&& 3.196 &  \\
& ${\frac{1}{2}}^+$ & 3.277 & 3.354 & 3.287 & 3.222\\
& ${\frac{3}{2}}^+$ & 3.268 & 3.335 & 3.282 & 3.217 & \\
& ${\frac{5}{2}}^+$ & 3.256 & 3.308 & 3.286 & 3.187 &&&&& 3.203 \\
& ${\frac{7}{2}}^+$ & 3.241 & 3.276 & 3.283 & 3.237 &&&&& 3.206 \\
\noalign{\smallskip}\hline
$2D$ & ${\frac{3}{2}}^+$ & 3.616 & 3.709 & 3.627 \\
& ${\frac{5}{2}}^+$ & 3.604 & 3.680 & 3.626 &&&\\
& ${\frac{1}{2}}^+$ & 3.629 & 3.741 & 3.623  \\
& ${\frac{3}{2}}^+$ & 3.620 & 3.719 & 3.613 \\
& ${\frac{5}{2}}^+$ & 3.608 & 3.691 & 3.614 \\
& ${\frac{7}{2}}^+$ & 3.594 & 3.656 & 3.611 \\
\noalign{\smallskip}\hline
$3D$ & ${\frac{3}{2}}^+$ & 3.958  \\
& ${\frac{5}{2}}^+$ & 3.947 \\
& ${\frac{1}{2}}^+$ & 3.971 \\
& ${\frac{3}{2}}^+$ & 3.962 \\
& ${\frac{5}{2}}^+$ & 3.951 \\
& ${\frac{7}{2}}^+$ & 3.938 \\
\noalign{\smallskip}\hline
$4D$ & ${\frac{3}{2}}^+$ & 4.292 \\
& ${\frac{5}{2}}^+$ & 4.282 \\
& ${\frac{1}{2}}^+$ & 4.303 \\
& ${\frac{3}{2}}^+$ & 4.296 \\
& ${\frac{5}{2}}^+$ & 4.285 \\
& ${\frac{7}{2}}^+$ & 4.273 \\
\noalign{\smallskip}\hline
$5D$ & ${\frac{3}{2}}^+$ & 4.617 \\
& ${\frac{5}{2}}^+$ & 4.608 \\
& ${\frac{1}{2}}^+$ & 4.628 \\
& ${\frac{3}{2}}^+$ & 4.621 \\
& ${\frac{5}{2}}^+$ & 4.611  \\
& ${\frac{7}{2}}^+$ & 4.599 \\
\noalign{\smallskip}\hline
\end{tabular}
\end{table}

\begin{table}
\addtocounter{table}{-1}
\caption{Predicted masses of the orbital excited states of $\Omega{_{c}^0}$ baryon ($F$, $G$ and $H$ wave) (in GeV).}
\label{tab:5}       
\begin{tabular}{lllllllllllllll}
\hline\noalign{\smallskip}
State & $J^P$ & Present & \cite{Shah2016epja} & \cite{Ebert2011} \\
\noalign{\smallskip}\hline\noalign{\smallskip}
$1F$ & ${\frac{5}{2}}^-$ & 3.536 & 3.602 & 3.522 \\
& ${\frac{7}{2}}^-$ & 3.519 & 3.565 & 3.498 \\
& ${\frac{3}{2}}^-$ & 3.555 & 3.643 & 3.533 \\
& ${\frac{5}{2}}^-$ & 3.541 & 3.613 & 3.515 \\
& ${\frac{7}{2}}^-$ & 3.524 & 3.577 & 3.514 \\
& ${\frac{9}{2}}^-$ & 3.503 & 3.532 & 3.485 \\
\noalign{\smallskip}\hline
$2F$ & ${\frac{5}{2}}^-$ & 3.878 \\
& ${\frac{7}{2}}^-$ & 3.863 \\
& ${\frac{3}{2}}^-$ & 3.895 \\
& ${\frac{5}{2}}^-$ & 3.883 \\
& ${\frac{7}{2}}^-$ & 3.868 \\
& ${\frac{9}{2}}^-$ & 3.850  \\
\noalign{\smallskip}\hline
$3F$ & ${\frac{5}{2}}^-$ & 4.213 \\
& ${\frac{7}{2}}^-$ & 4.199 \\
& ${\frac{3}{2}}^-$ & 4.229 \\
& ${\frac{5}{2}}^-$ & 4.218 \\
& ${\frac{7}{2}}^-$ & 4.204 \\
& ${\frac{9}{2}}^-$ & 4.186 \\
\noalign{\smallskip}\hline
$4F$ & ${\frac{5}{2}}^-$ & 4.541 \\
& ${\frac{7}{2}}^-$ & 4.527 \\
& ${\frac{3}{2}}^-$ & 4.556 \\
& ${\frac{5}{2}}^-$ & 4.545 \\
& ${\frac{7}{2}}^-$ & 4.531 \\
& ${\frac{9}{2}}^-$ & 4.515 \\
\noalign{\smallskip}\hline
$1G$ & ${\frac{7}{2}}^+$ & 3.800 &  & 3.721 \\
& ${\frac{9}{2}}^+$ & 3.777 &  & 3.685 \\
& ${\frac{5}{2}}^+$ & 3.824 &  & 3.739 \\
& ${\frac{7}{2}}^+$ & 3.806 &  & 3.707\\
& ${\frac{9}{2}}^+$ & 3.783 &  & 3.705  \\
& ${\frac{11}{2}}^+$ & 3.756 &  & 3.665\\
\noalign{\smallskip}\hline
$2G$ & ${\frac{7}{2}}^+$ & 4.136  \\
& ${\frac{9}{2}}^+$ & 4.115 \\
& ${\frac{5}{2}}^+$ & 4.158 \\
& ${\frac{7}{2}}^+$ & 4.141 \\
& ${\frac{9}{2}}^+$ & 4.120  \\
& ${\frac{11}{2}}^+$ & 4.096 \\
\noalign{\smallskip}\hline
$3G$ & ${\frac{7}{2}}^+$ & 4.461 \\
& ${\frac{9}{2}}^+$ & 4.445 \\
& ${\frac{5}{2}}^+$ & 4.478 \\
& ${\frac{7}{2}}^+$ & 4.465 \\
& ${\frac{9}{2}}^+$ & 4.449  \\
& ${\frac{11}{2}}^+$ & 4.431  \\
\noalign{\smallskip}\hline
$1H$ & ${\frac{9}{2}}^-$ & 4.057  \\
& ${\frac{11}{2}}^-$ & 4.029 \\
& ${\frac{7}{2}}^-$ & 4.087 \\
& ${\frac{9}{2}}^-$ & 4.063 \\
& ${\frac{11}{2}}^-$ & 4.035 \\
& ${\frac{13}{2}}^-$ & 4.004 \\
\noalign{\smallskip}\hline
$2H$ & ${\frac{9}{2}}^-$ & 4.386 \\
& ${\frac{11}{2}}^-$ & 4.362 \\
& ${\frac{7}{2}}^-$ & 4.412 \\
& ${\frac{9}{2}}^-$ & 4.391 \\
& ${\frac{11}{2}}^-$ & 4.367 \\
& ${\frac{13}{2}}^-$ & 4.339 \\
\noalign{\smallskip}\hline
\end{tabular}
\end{table}

\section{Methodology}
\label{sec:2}

Usually the relativistic and the non-relativistic treatments of quantum mechanics are used to study the spectroscopy of heavy and light flavored hadrons. Especially when dealing with a three quarks system (baryon), the problem is become more challenging. To describe the relative motion of the constituent quarks inside the baryon the Jacobi coordinates ($\vec{\rho}$ and $\vec{\lambda}$) are used. Which are given by \cite{Richard1992,Giannini2015},

\begin{equation}
\label{eqn:1} 
\vec{\rho} = \frac{(\vec{r_1}-\vec{r_2})}{\sqrt2} \hspace{0.5cm} and \hspace{0.5cm} \vec{\lambda}=\frac{m_1\vec{r_1}+m_2\vec{r_2}-(m_1+m_2)\vec{r_3}} {\sqrt {m_1^2+m_2^2+(m_1+m_2)^2}}.
\end{equation}

\noindent where $\vec{r}_i$ and $m_i$ $(i=1,2,3)$ denote the spatial and the mass coordinate of the $i-$th constitute quark. We use hypercentral Constituent Quark Model (hCQM), which is well established for the study of light, heavy-light and heavy flavored baryons \cite{Shah2016epja,Shah2019,Ghalenovia2014,Salehi2011,Santopinto2005}. In hCQM, the Hamiltonian for a  baryonic system is,

\begin{equation}
\label{eqn:2} 
H=\frac{P{_{x}^2}}{2m} + V(x),
\end{equation}

\noindent where $x = (\vec{\rho}, \vec{\lambda})$ is the radial hypercentral coordinate of the three quarks system written in a six-dimensional space and the hyperangle $\xi= arctan \left( \frac{\rho^2}{\lambda^2} \right)$. $ m = \frac{2m_\rho m_\lambda}{m_\rho+m_\lambda} $ gives the reduced mass of the three quarks \cite{Richard1992}. The reduced mass $m_\rho$ and $m_\lambda$ corresponding to the Jacobi coordinates $\vec{\rho}$ and $\vec{\lambda}$ are,  

\begin{equation}
\label{eqn:3} 
m_\rho = \frac{2m_1m_2}{(m_1+m_2)}  \hspace{0.5cm} and \hspace{0.5cm} m_\lambda =\frac{2m_3(m{_{1}^2}+m{_{1}^2}+m_1m_2)}{(m_1+m_2)(m_1+m_2+m_3)}.
\end{equation}

Here, $m_1$, $m_2$ and $m_3$ are the constituent quark masses. The isodoublet  of the $\Xi{_c}$ baryons has a different quark constitution:  $\Xi{_{c}^0}$ $(dsc)$, $\Xi{_{c}^+}$ $(usc)$ and for the singlet $\Omega{_{c}^0}$ baryon the constituent quarks are $s$, $s$ and $c$. To calculate the masses of  isodublet $\Xi{_c}$ baryons separately, we consider the unequal masses of light quarks $(u,d)$ as: $m_u$ = 0.338 GeV, $m_d$ = 0.35 GeV and $m_s$ = 0.5 GeV, and the charm ($c$) quark mass $m_c$ = 1.275 GeV.\\

The potential $V(x)$ appearing into the Eq. (\ref{eqn:2}) is the non-relativistic hypercentral potential and is divided into spin independent $V_{SI}(x)$ and spin dependent $V_{SD}(x)$ part of the potential. We introduce $V_{SI}(x)$ as,

\begin{equation}
\label{eqn:4} 
V_{SI}(x)=V_V(x)+V_S(x)+V_I(x).
\end{equation}
  
\noindent Here, the first order correction $V_I(x) = - C_F C_A \frac{\alpha{_{S}^2}}{4x^2}$; is written in the form of Casimir charges, with $C_F = \frac{2}{3} $ and $ C_A = 3 $ in the fundamental and adjoint representation respectively \cite{Koma2006}. The vector (Coulomb) potential is, 

\begin{equation}
\label{eqn:5} 
V_V(x) = -\frac{2}{3}\frac{\alpha_s}{x} 
\end{equation}

\noindent and it gives the interaction between quarks in a baryonic system. $\alpha_s$ is the strong running coupling constant and is,

\begin{equation}
\label{eqn:6} 
\alpha_s = \frac{\alpha_s(\mu_0)}{1+\left( \frac{33-2n_f}{12\pi}\right)\alpha_s(\mu_0)ln\left(\frac{m_1+m_2+m_3}{\mu_{0}}\right)}.
\end{equation}

\noindent Here, we take $\alpha_s(\mu_0=1GeV) = 0.6$ and $n_f$ = 3 gives the number of active quark flavors contributing in quark-gloun loops. We choose a screened potential as a scalar (confining) potential,

\begin{equation}
\label{eqn:7} 
V_{S}(x) = a \left(\frac{1-e^{-\mu x}}{\mu}\right)
\end{equation}

\noindent where $a (\approx 0.1)$ is the string tension and $\mu= 0.04$ GeV is the flavor independent parameter called screening factor. Using $a$, $\mu(=$ 0.04 GeV) and quark masses, we fix the ground state $(1S)$ mass of these baryons to the known experimental value (PDG) \cite{Tanabashi2018}. If the value of $\mu=0.04$ GeV is change  by $\pm$ 10\% the changes in the masses of $2S$, $3S$, $4S$, $5S$ and $6S$ states are about 0.30\%, 0.53\%, 0.75\%, 0.93\% and 1.1\% respectively. For $x \ll \frac{1}{\mu}$, the screened potential is behaving like a linear potential $ax$ and for $ x \gg \frac{1}{\mu} $ it becomes a constant $ \frac{a}{\mu} $. The mass spectroscopy resulting from a screened potential is interesting: it predicts a smaller masses of the higher excited states with respect to a linear potential (see results in Ref. \cite{Shah2016epja} for $\nu=$ 1 and, more detailed studies in Ref. \cite{JZWang2019}). By taking an unequal $u$ and $d$ quarks masses the mass spectra of $\Xi_c^0$, $\Xi_c^+$ and $\Omega_c^0$ baryons are calculated in this work.\\

Using these parameters we compute the spin-averaged mass of the strange singly charmed baryons. In order to calculate the masses of $nL$ with different $J$ values the spin dependent potential $V_{SD}(x)$ is employed \cite{Voloshin2008,Bijker1994,Bijker1998},

\begin{equation}
\label{eqn:8} 
V_{SD}(x) = V_{SS}(x)(\vec{S_\rho}\cdot \vec{S_\lambda}) + V_{\gamma S}(x)(\vec{L} \cdot \vec{S}) + V_{T}(x) \left[S^2 - \frac{3(\vec S \cdot \vec x)(\vec S \cdot \vec x)}{x^2}\right].
\end{equation}

\noindent The spin-spin interaction $V_{SS}(x)$ gives a hyperfine mass splitting, and the spin-orbit $V_{\gamma S}(x)$ and spin-tensor $V_{T}(x)$ interaction terms gives a fine structure of the given baryonic states (for details see Ref. \cite{Shah2016epja}).\\ 

\noindent The six-dimensional hyperradial Schr\"{o}dinger equation is,
\begin{equation}
\label{eqn:14}  
\left[\frac{1}{2m}\left(-\frac{d^2}{dr^2}+\frac{\frac{15}{4}+\gamma(\gamma+4)}{r^2}\right)+ V(x)\right]\phi_{\gamma}(x) = E_B \phi_{\gamma}(x),
\end{equation}

\noindent solves numerically using mathematica notebook \cite{Lucha1999}. $\phi_{\gamma}(x)$ is the hypercentral radial function and $E_B$ gives the binding energy of the baryonic states. Our computed mass spectra of strange singly charmed baryons are presented in Tables \ref{tab:2}-\ref{tab:5}. We discussed them into the next section.

%
\begin{figure*}
\resizebox{1.00\textwidth}{!}{%
  \includegraphics{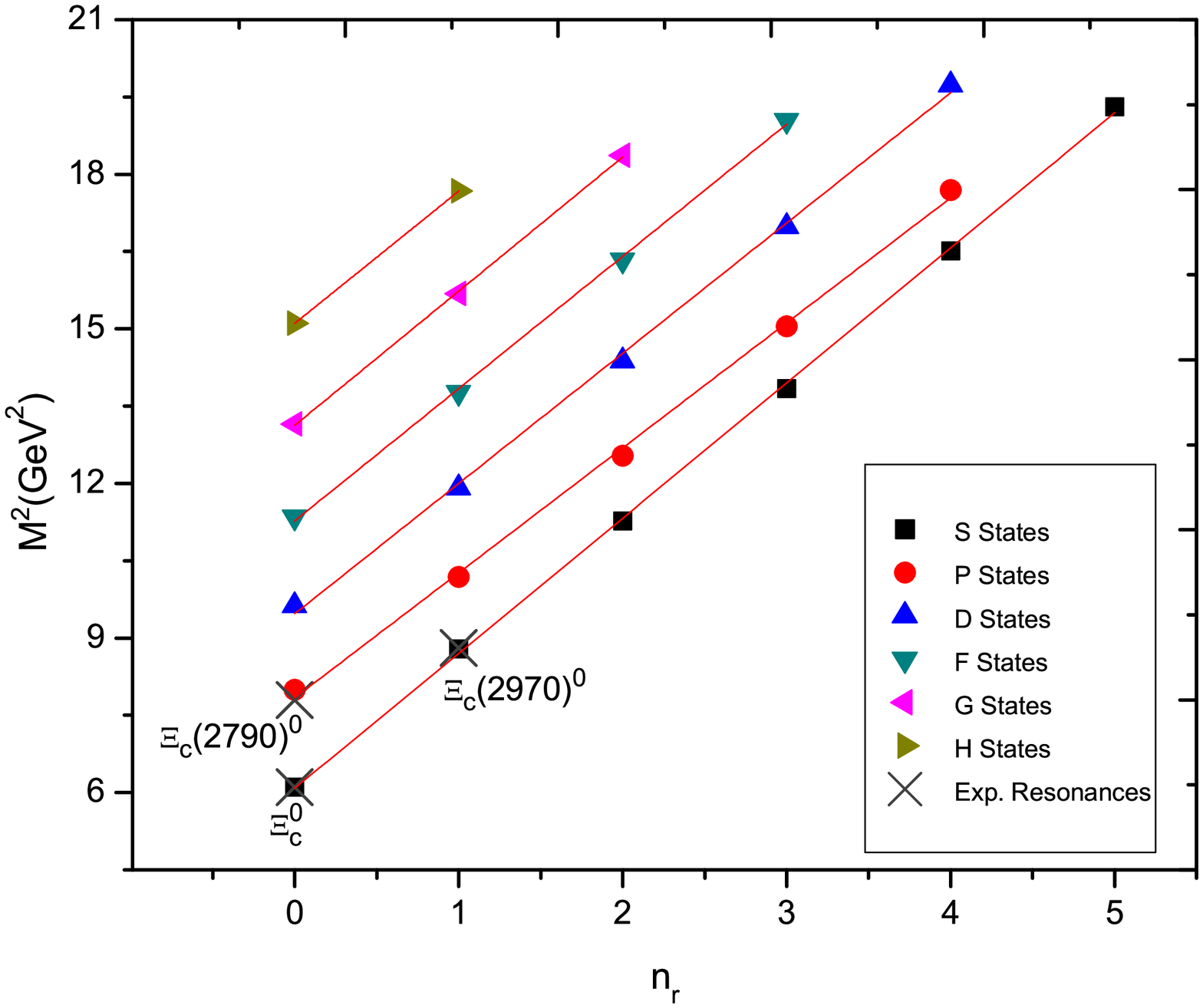}
}
\caption{The $(M^2 \rightarrow n_r)$ Regge trajectory of $\Xi{_{c}^0}$ baryon.}
\label{fig:1}       
\end{figure*}

\begin{figure*}
\resizebox{1.00\textwidth}{!}{%
  \includegraphics{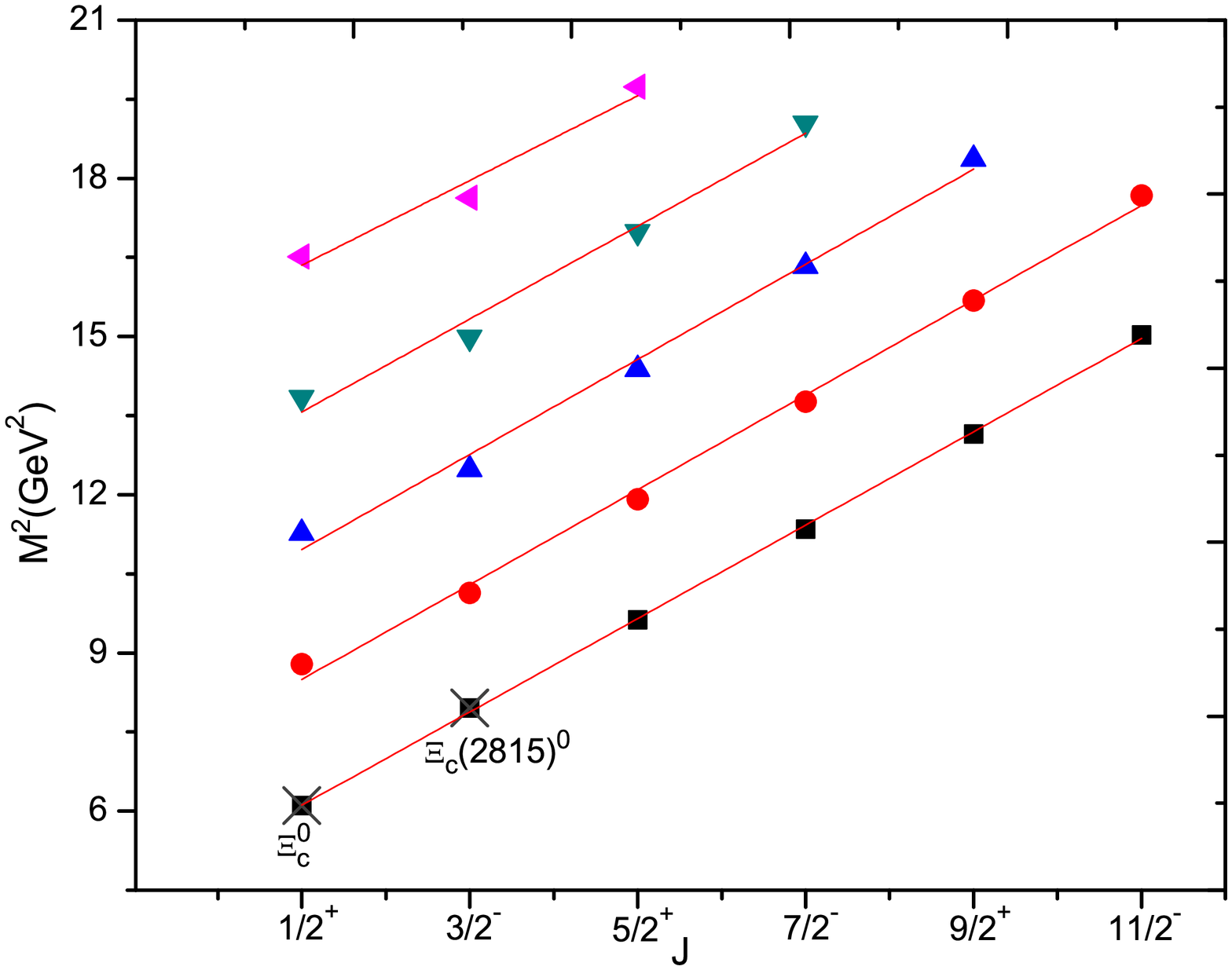}
   \includegraphics{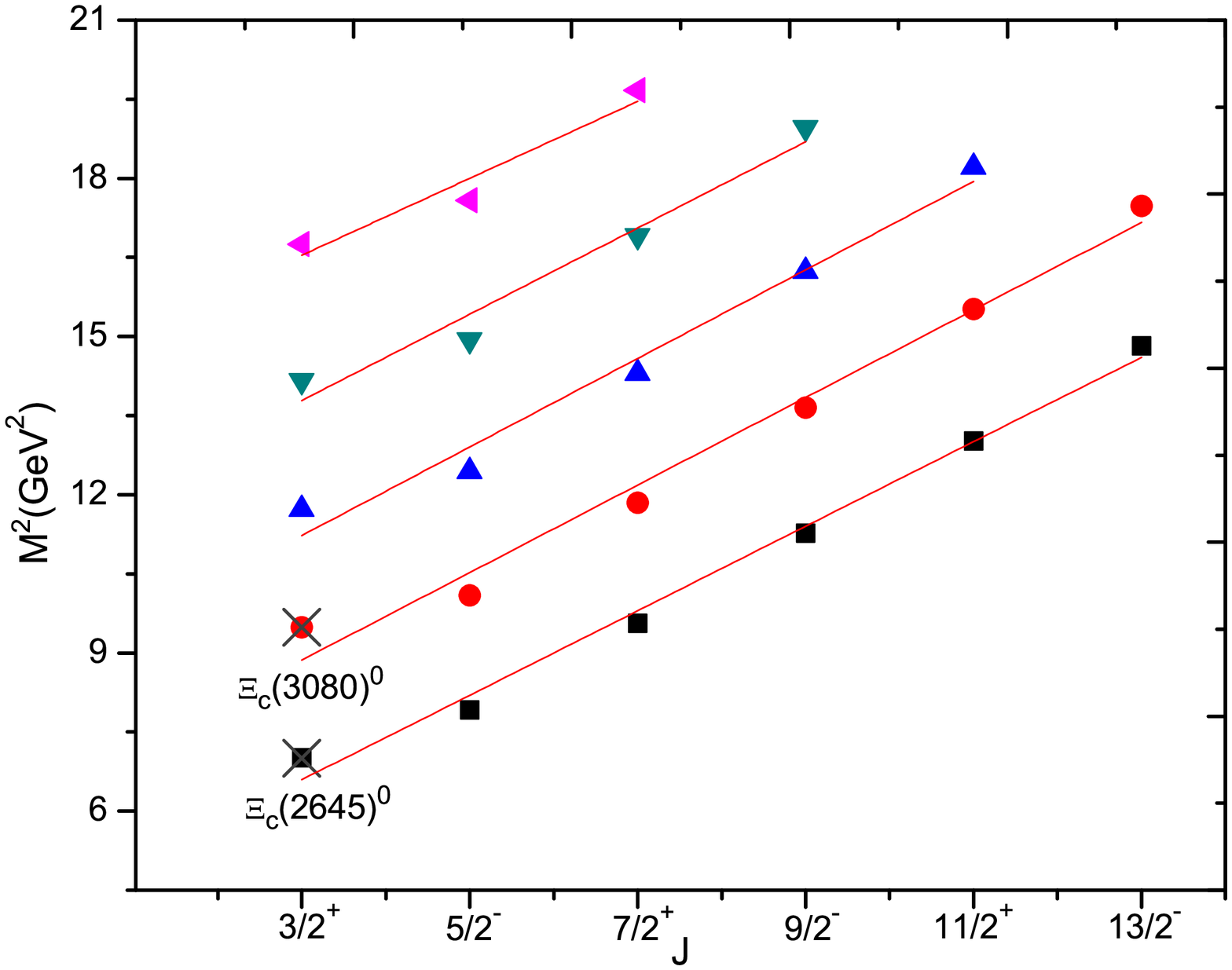}
}
\caption{The $(M^2 \rightarrow J)$ Regge trajectories of $\Xi{_{c}^0}$ baryon with the natural parity (left) and the unnatural parity (right).}
\label{fig:2}       
\end{figure*}

\begin{figure*}
\resizebox{1.00\textwidth}{!}{%
  \includegraphics{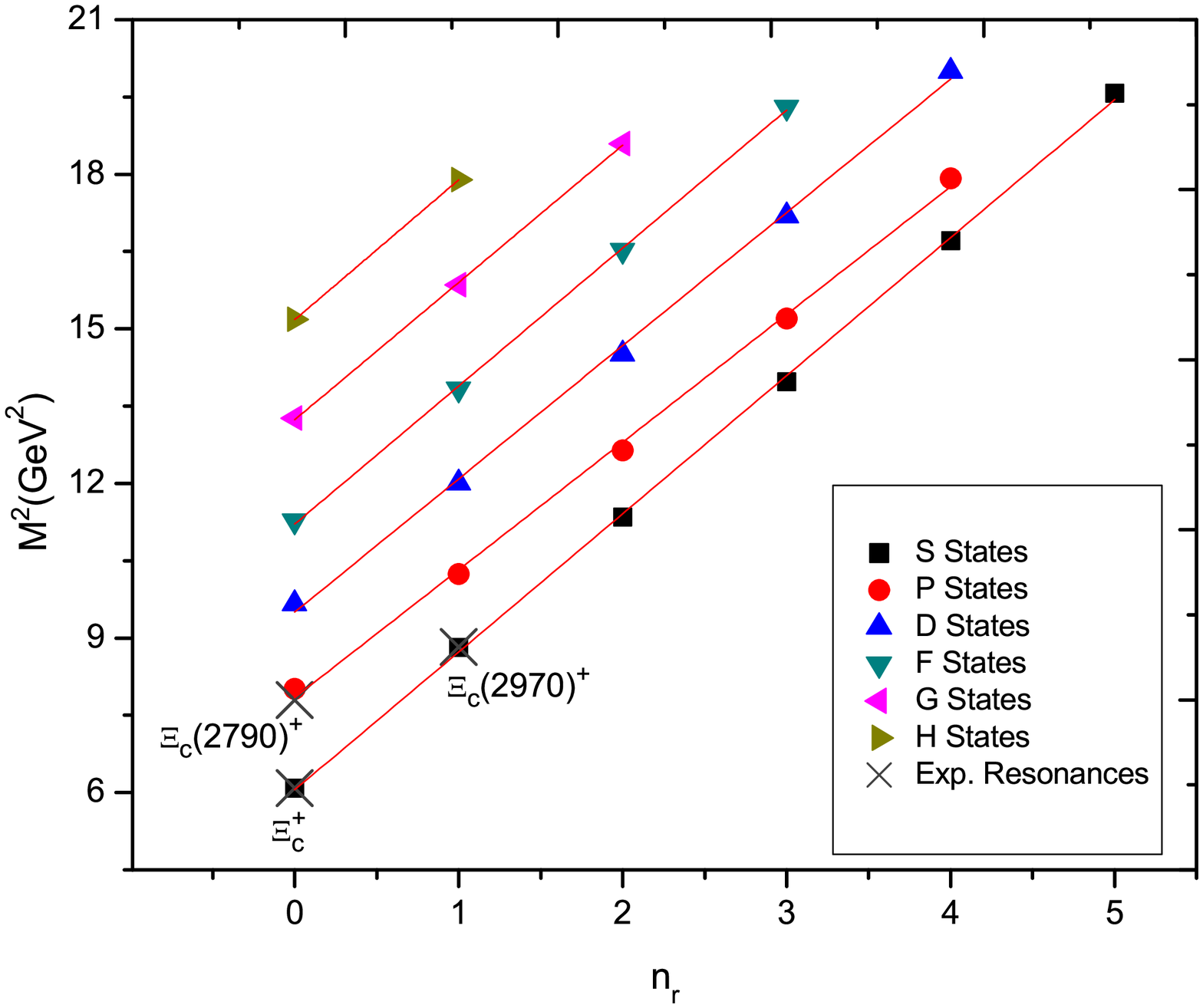}
}
\caption{The $(M^2 \rightarrow n_r)$ Regge trajectory of $\Xi{_{c}^+}$ baryon.}
\label{fig:3}       
\end{figure*}

\begin{figure*}
\resizebox{1.00\textwidth}{!}{%
  \includegraphics{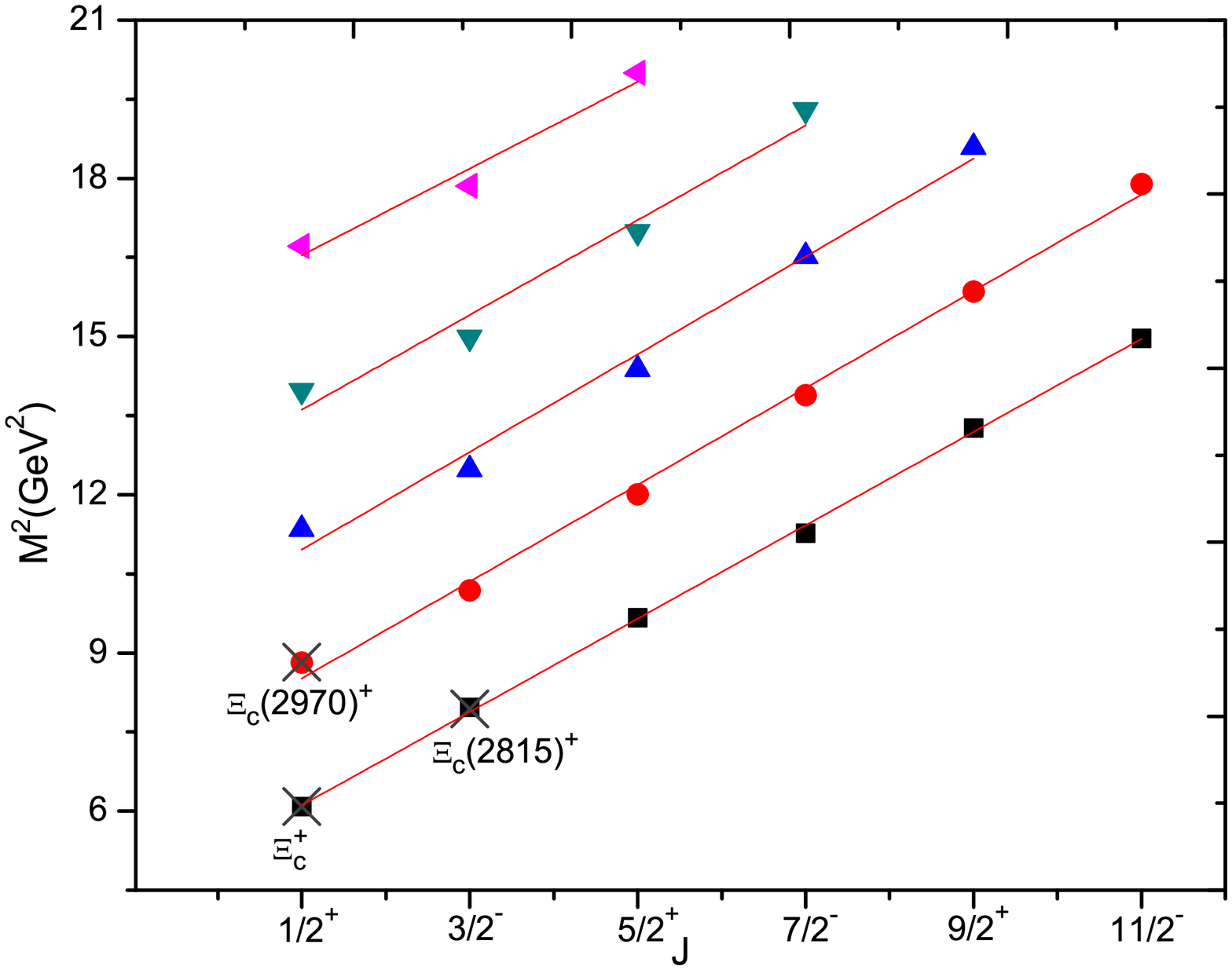}
  \includegraphics{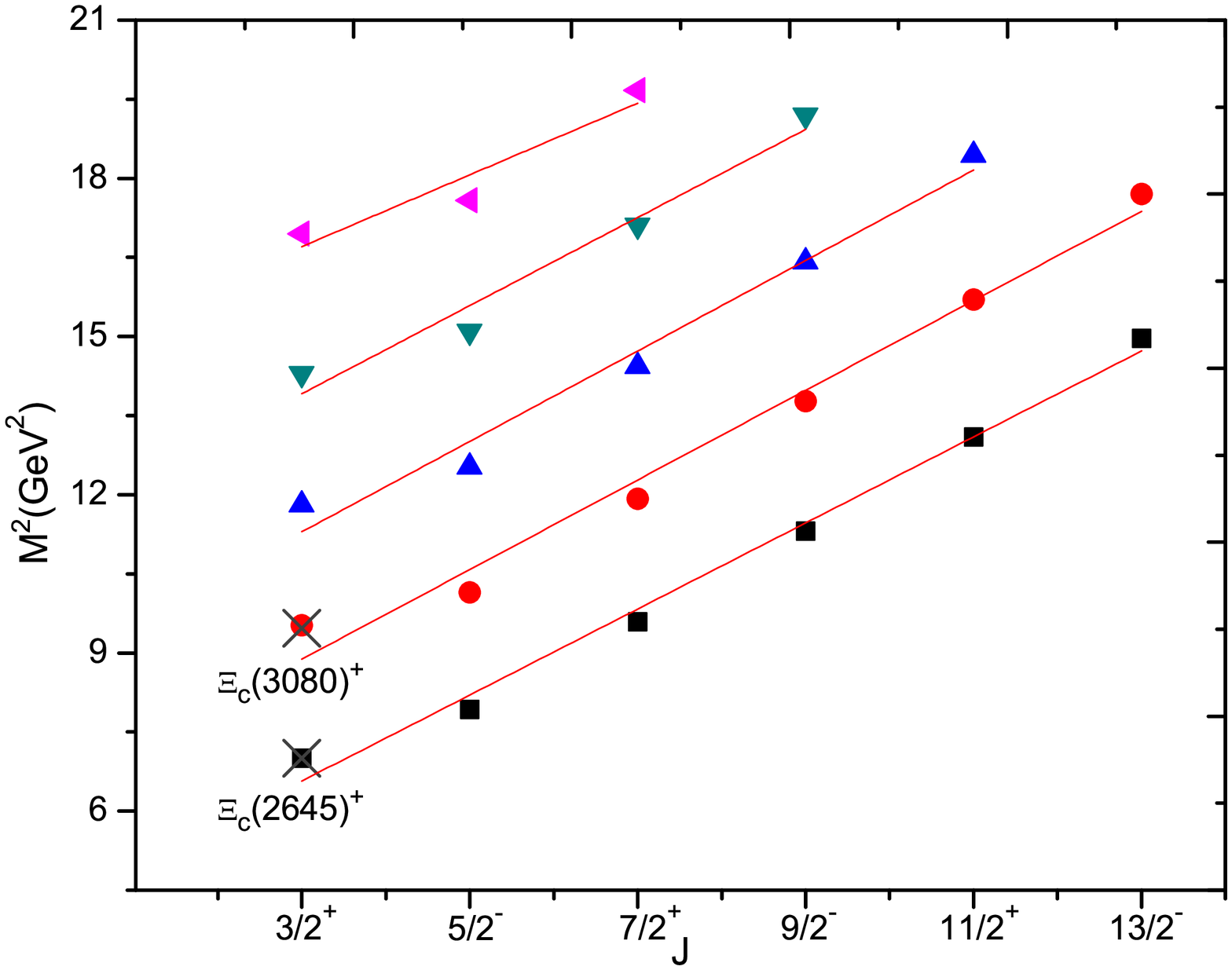}
}
\caption{The $(M^2 \rightarrow J)$ Regge trajectories of $\Xi{_{c}^+}$ baryon with the natural parity (left) and the unnatural parity (right).}
\label{fig:4}       
\end{figure*}  

\begin{figure*}
\resizebox{1.00\textwidth}{!}{%
  \includegraphics{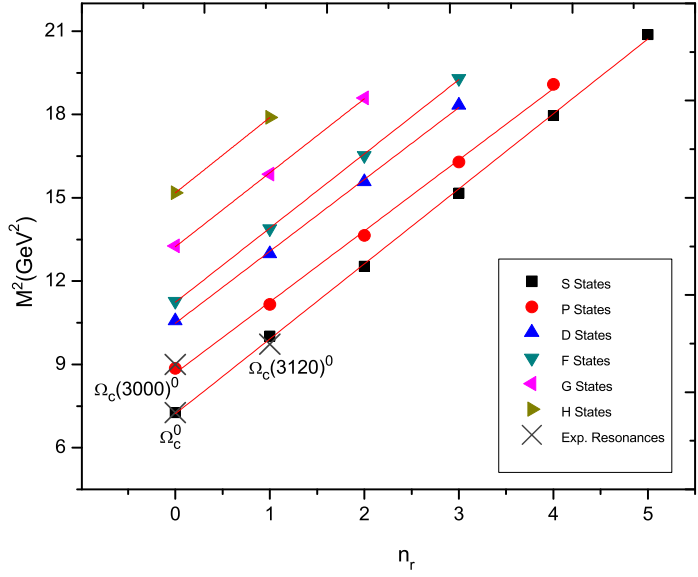}
}
\caption{The $(M^2 \rightarrow n_r)$ Regge trajectory of $\Omega{_{c}^0}$ baryon.}
\label{fig:5}       
\end{figure*}

\begin{figure*}
\resizebox{1.00\textwidth}{!}{%
  \includegraphics{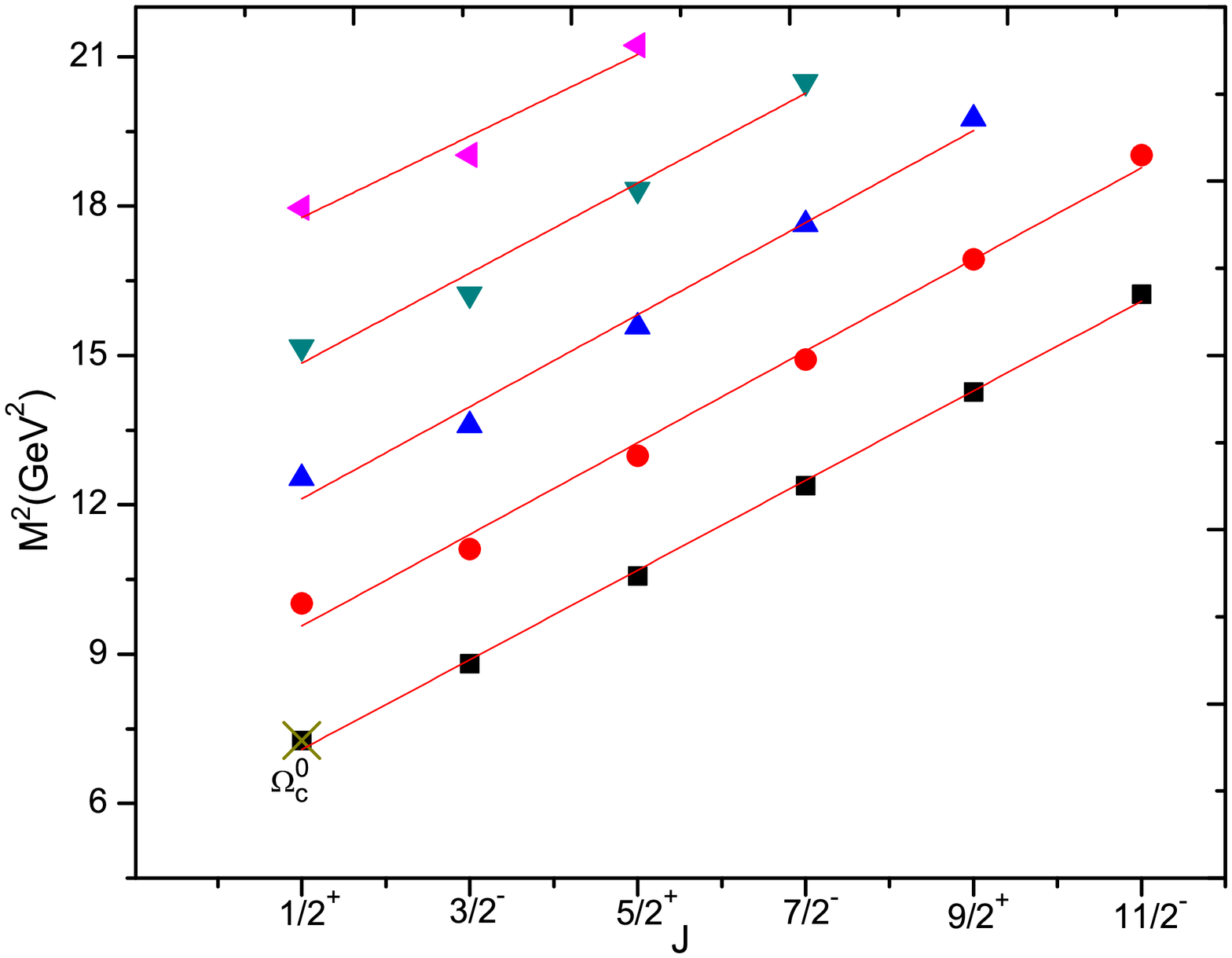}
   \includegraphics{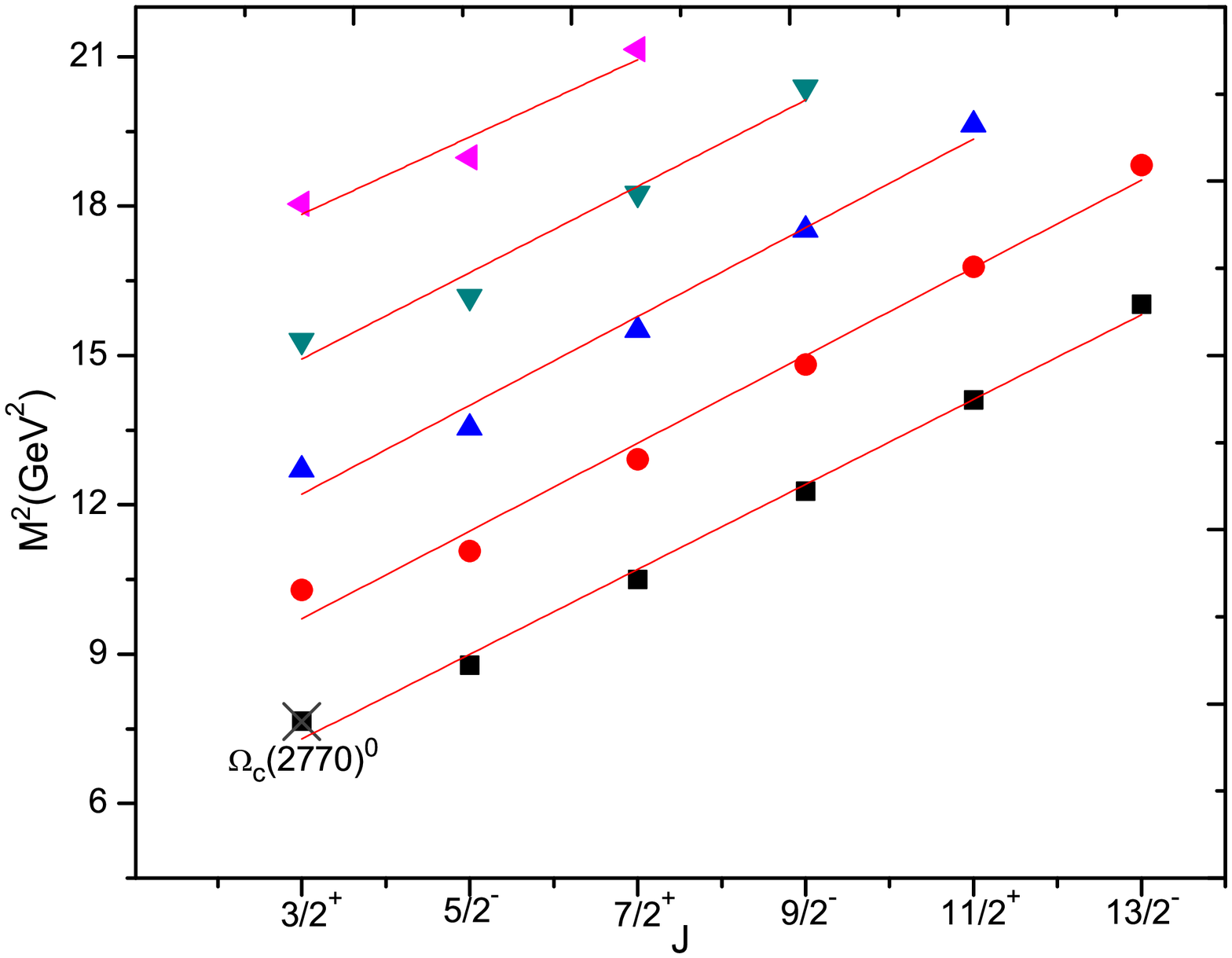}
}
\caption{The $(M^2 \rightarrow J)$ Regge trajectories of $\Omega{_{c}^0}$ baryon with the natural parity (left) and the unnatural parity (right).}
\label{fig:6}       
\end{figure*}

\section{Mass Spectra and Regge Trajectories}
\label{sec:3}

The mass spectrum of strange singly charmed baryons are calculated in the non-relativistic framework of hypercentral Constituent Quark Model (hCQM). The screened potential is used as a scalar potential with the first order correction. Our calculated masses of the radial excited states ($2S$-$6S$) are presented in Table \ref{tab:2} and the orbital excited states ($1P$-$5P$, $1D$-$5D$, $1F$-$4F$, $1G$-$3G$ and, $1H$ and $2H$) are listed in Tables \ref{tab:3}-\ref{tab:5}. Our results are compared with other theoretical predictions and the experimental observations where available. The obtained masses are used to draw Regge trajectories in the $(n_r, M^2)$  and $(J, M^2)$ planes. They are used to determine the possible quantum number of the particular hadronic states \cite{Guo2008}. The $S$-states with $J^P$ = ${\frac{1}{2}}^+$ and the $P$, $D$, $F$, $G$, $H$-states for the $J^P$ values ${\frac{1}{2}}^-$, ${\frac{5}{2}}^+$, ${\frac{7}{2}}^-$, ${\frac{9}{2}}^+$, ${\frac{11}{2}}^-$ respectively; are plotted in the $(n_r, M^2)$ plane. Using the equation,    

\begin{equation}
\label{eqn:15} 
n_r = \beta M^2 + \beta_0
\end{equation}

\noindent where $n_r$ = $n-1$ and, $\beta$ and $\beta_0$ are the fitted slopes and the intercepts respectively. In the $(J, M^2)$ plane, the Regge trajectories are plotted with natural $P= (-1)^{J-\frac{1}{2}}$ and the unnatural $P= (-1)^{J+\frac{1}{2}}$ parities. Therefore, the states with $J^P$ values ${\frac{1}{2}}^+$, ${\frac{3}{2}}^-$, ${\frac{5}{2}}^+$, ${\frac{7}{2}}^-$, ${\frac{9}{2}}^+$ and ${\frac{11}{2}}^-$ are here available with the natural parities; and the states ${\frac{3}{2}}^+$, ${\frac{5}{2}}^-$, ${\frac{7}{2}}^+$, ${\frac{9}{2}}^-$, ${\frac{11}{2}}^+$ and ${\frac{13}{2}}^-$ are for the unnatural parities. Using the equation,          

\begin{equation}
\label{eqn:16}  
J=\alpha M^2+\alpha_0
\end{equation}

\noindent where $\alpha$ and $\alpha_0$ are the fitted slopes and the intercepts respectively. The Regge trajectories of the strange singly charmed baryons are shown in Figs. \ref{fig:1}-\ref{fig:6} plotted in the $(n_r, M^2)$  and $(J, M^2)$ planes using the above equations. On the Regge line our calculated masses are presented and the available experimental observations are indicated by a cross sign with a possible quantum number.

\subsection{$\Xi{_c}$ states}

The spin-parity of the ground $(1S)$ state of $\Xi{_c^+}$ baryon and its neutral isospin partner $\Xi{_c^0}$ were confirmed experimentally. For the predictions of the excited radial and orbital states of isodoublet $\Xi_c$ baryons, we fixed their masses of $\Xi{_c^+}$: 2.46787 $\pm$ 0.0003 GeV and $\Xi{_c^0}$: 2.47087${_{-0.00031}^{+0.00028}}$ GeV for $J^P$ = ${\frac{1}{2}}^+$; and for the $J^P$ = ${\frac{3}{2}}^+$ the $\Xi{_c^+}$: 2.64553 $\pm$ 0.00031 GeV and $\Xi{_c^0}$: 2.64632 $\pm$ 0.00031 GeV from \cite{Tanabashi2018}. The isodoublet of $\Xi_c$ baryons ($\Xi_c^0(dsc)$ and $\Xi_c^+(usc)$) has different quark constituents. The authors of Refs. \cite{BChen2017,Roberts2008,Ebert2011,Ebert2008,BChen2015,Yamaguchi2015,Valcarce2008} consider equal mass of light quarks $u$ and $d$. So they can?t distinguish the isodoublet of $\Xi_c$ baryons and calculate their excited states masses. In this work we use $m_u=$ 0.338 GeV and $m_d=$ 0.350 GeV are taken from Ref. \cite{Shah2016epja,} and we calculate the mass spectra of $\Xi_c^0(dsc)$ and $\Xi_c^+(usc)$ baryons separately (see in Tables \ref{tab:2}-\ref{tab:4}). So we can compare our results with the predictions of Refs. \cite{BChen2017,Roberts2008,Ebert2011,Ebert2008,BChen2015,Yamaguchi2015,Valcarce2008} with either $\Xi_c^0$ baryon or either $\Xi_c^+$ baryon. In this work, our results of radial and orbital excited states of $\Xi_c^0$ baryon are compared with the predictions of Refs.  \cite{BChen2017,Roberts2008,Ebert2011,Ebert2008,BChen2015,Yamaguchi2015,Valcarce2008}.\\

In 2016, the Belle Collaboration \cite{Yelton2016} studied the decay of an isodoublet of $\Xi_{c}(2790)$ and $\Xi_{c}(2815)$ into $\Xi{_c^0}$ and $\Xi{_c^+}$ final states. They identified $\Xi{_c(2790)}$ and $\Xi{_c(2815)}$ states with the total spin ${\frac{1}{2}}^-$ and ${\frac{3}{2}}^-$ respectively. According to the quark mass hierarchy the down $(d)$ quark is heavier than up $(u)$ quark. So the $\Xi{_c^0}$ are expected to be heavier than $\Xi{_c^+}$ masses. Ref. \cite{Yelton2016} gives a negative isospin mass splittings of the $\Xi{_c}$ isodoublets; $M(\Xi{_c(2645)^+}$ - $\Xi{_c(2645)^0})$, $M(\Xi{_c(2790)^+}$ - $\Xi{_c(2790)^0})$, $M(\Xi{_c(2815)^+}$ - $\Xi{_c(2815)^0})$, $M(\Xi{_c(2970)^+}$ - $\Xi{_c(2970)^0})$ and $M(\Xi{_c(3080)^+}$ - $\Xi{_c(3080)^0})$. In the present study, the isospin mass splittings are positive, i.e. $\Xi{_c^+} > \Xi{_c^0}$ (about isospin mass splitting see Ref. \cite{Silvestre-Brac2003}).\\      

Also, the Ref. \cite{Yelton2016} observed an isodoublet $\Xi{_c(2970)}$ resonances, but still their $J^P$ values are not found. Their world average masses are 2.9707 $\pm$ 0.0022 GeV and 2.968 $\pm$ 0.0026 GeV \cite{Tanabashi2018}. These are in accordance with our predictions of $2S$ state with $J^P$ = ${\frac{1}{2}}^+$. The world average masses of the isodoublet of $\Xi{_c(3080)}$ are 3.0772 $\pm$ 0.0004 GeV and 3.0799 $\pm$ 0.0014 GeV \cite{Tanabashi2018}. Which are in agreement with our predictions of a $2S$ state with $J^P$ = ${\frac{3}{2}}^+$ (see in Table \ref{tab:2}). The BABAR Collaboration \cite{Aubert20081} reported a state $\Xi{_c(3123)^+}$ of mass 3.1229 $\pm$ 0.0013(stat)  $\pm$ 0.0003(syst) GeV. This is exactly matched by our prediction of the second orbital excitation of $\Xi{_c^+}$ with $J^P$ = ${\frac{3}{2}}^+$; for the spin $S$ = ${\frac{1}{2}}$ (see in Table \ref{tab:4}).\\

For the radial excitation of $\Xi{_c^0}$ baryon, with $J^P$ = ${\frac{1}{2}}^+$ and $J^P$ = ${\frac{3}{2}}^+$, our predictions are larger than Refs. \cite{BChen2017,Ebert2011,Ebert2008,Yamaguchi2015,Valcarce2008}. For the $P$ and $D$-states, our results are accordance with the predictions of Refs. \cite{BChen2017,Ebert2011,Ebert2008,BChen2015}. And for $1F$ to $4F$, $1G$-$2G$ and $1H$ states our predictions are larger than Ref. \cite{Ebert2011} by mass differences between 77 MeV and 262 MeV (see in Table \ref{tab:3}). For the isodoublet $\Xi{_c}$ baryons our results are in accordance with the predictions of Z. Shah \textit{et al.} \cite{Shah2016epja}. The other theoretical and phenomenological studies consider same light quark masses, so here our predictions are compared only with $\Xi{_c^0}$ baryon. Our calculated masses of the radial excited states of $\Xi{_c^+}$ baryon are presented in Table \ref{tab:2} and their orbital excited states are listed in Table \ref{tab:4}.\\

The Regge trajectories of the isodoublet $\Xi{_c}$ baryons are shown in Figs. \ref{fig:1}-\ref{fig:4}. In both the planes $(n_r, M^2)$  and $(J, M^2)$ their squared masses are fit nicely on a Regge lines. Regge trajectories confirmed the isodoublet states of $\Xi{_c(2790)}$ and $\Xi{_c(2815)}$ as a first orbital excited states with spin-parity ${\frac{1}{2}}^-$ and ${\frac{3}{2}}^-$ respectively. They predict an isodoublet of $\Xi{_c(2970)}$ and $\Xi{_c(3080)}$ as a $2S$-states with spin-parity ${\frac{1}{2}}^+$ and ${\frac{3}{2}}^+$ respectively. In the present study, the Regge trajectories are enabled for the predictions of the experimental observed states $\Xi{_c(2930)}$ and $\Xi{_c(3055)^+}$.   

\subsection{$\Omega{_c^0}$ states}

Experimentally, only the ground states of $\Omega{_c^0}$ baryon have been observed with its quantum number. PDG \cite{Tanabashi2018} fitted their masses: 2.6952 $\pm$ 0.0017 GeV for $J^P$ = ${\frac{1}{2}}^+$; and 2.7659 $\pm$ 0.002 GeV  for $J^P$ = ${\frac{3}{2}}^+$. The $\Omega{_c}(2770)^0$ was observed by BABAR Collaboration \cite{Aubert2006} in the decay of $\Omega{_c}(2770)^0 \rightarrow \Omega{_{c}^{0}} + \gamma$. Recently, the LHCb Collaboration \cite{Aaij2017} observed a number of excited states of $\Omega{_c^0}$ baryon: $\Omega{_c}(3000)^0$, $\Omega{_c}(3050)^0$, $\Omega{_c}(3066)^0$, $\Omega{_c}(3090)^0$ and $\Omega{_c}(3119)^0$. Their $J^P$ values are not confirmed yet.\\

We calculate the masses of radial and orbital excited states of $\Omega{_c^0}$ baryon. For that, first we fit their ground state masses from Ref. \cite{Tanabashi2018}. The LHCb \cite{Aaij2017} and the Belle Collaboration \cite{Yelton2018} reported a state $\Omega{_c}(3000)^0$. The PDG \cite{Tanabashi2018} confirmed its world average mass, 3.00041 $\pm$ 0.00022 GeV. This state is close to our prediction 2.976 GeV of $1P$-state of $J^P$ = ${\frac{1}{2}}^-$ with spin $S$ = ${\frac{1}{2}}$ (see in Table \ref{tab:5}). Also, they report a state $\Omega{_c}(3119)^0$ measured with the mass 3.1191 $\pm$ 0.0003(stat) $\pm$ 0.0009${_{-0.0005}^{+0.0003}}$(syst) GeV \cite{Aaij2017}. Such a state might be $2S$-state with $J^P$ = ${\frac{1}{2}}^+$ (see in Table \ref{tab:2}). For the radial excited states of $\Omega{_c^0}$ baryon, with $J^P$ = ${\frac{1}{2}}^+$ and $J^P$ = ${\frac{3}{2}}^+$, our results are in agreement with the predictions of Refs. \cite{Roberts2008,Ebert2011}. And, for the orbital excitations of $1P$ to $4P$, $1D$-$2D$, $1F$ and $1G$ states, our results are in accordance with the predictions of Refs. \cite{Yoshida2015,Roberts2008,Ebert2011,Ebert2008,Yamaguchi2015,Chen2015,Valcarce2008}. In the $\Omega{_c^0}$ baryon mass spectrum, the screening effect contributes to predictions of lower masses of the excited radial and orbital states with respect to the results obtained by Z. Shah \textit{et al.} \cite{Shah2016epja} using the linear potential.\\

The Regge trajectories of the $\Omega{_c^0}$ baryon are shown in Figs. \ref{fig:5}-\ref{fig:6}. In both the planes $(n_r, M^2)$  and $(J, M^2)$ the squared masses of its excited states are fit nicely on a Regge lines. The $\Omega{_c}(3000)^0$ is predicted as a first orbital excited state with $J^P$ = ${\frac{1}{2}}^-$ and the $\Omega{_c}(3119)^0$ may have $2S$-state with $J^P$ = ${\frac{1}{2}}^+$. In the present study, the Regge trajectories are enabled for the predictions of the experimental observed states $\Omega{_c}(3050)^0$, $\Omega{_c}(3066)^0$ and $\Omega{_c}(3090)^0$.    

\section{Properties}
\label{sec:4}

\begin{table}
\caption{The strong decay widths (in MeV) of $\Xi_c$ baryons.}
\label{tab:6}       
\scalebox{0.9}{
\begin{tabular}{lllllllllllllllllllllllllll}
\hline\noalign{\smallskip}
State & Decay Mode & HHChPT & HHChPT(2019) & \cite{Cheng20151} & \cite{Cheng2007} & \cite{JYKim2019} & \cite{Ivanov1999} & \cite{Tawfiq1998} & \cite{Albertus2005} & Exp. \cite{Yelton2016} \\
& & Present & PDG-2018 & & \\
\noalign{\smallskip}\hline\noalign{\smallskip}
$\Xi{_c(2645)^+}$ & $\Xi{_{c}^{+}} \pi^0$ & $0.97{_{-0.09}^{+0.05}}$ & $0.92{_{-0.09}^{+0.04}}$ \\
& $\Xi{_{c}^{0}} \pi^+$ & $1.41{_{-0.13}^{+0.07}}$  & $1.39{_{-0.13}^{+0.07}}$ \\
& Total & $2.38{_{-0.23}^{+0.11}}$ & $2.31{_{-0.22}^{+0.11}}$ & 2.8 & 2.7 & & 3.04 & 1.76 & 3.18 & 2.06\\
&&&& $\pm$ 0.4 & $\pm$ 0.2 && $\pm$ 0.37 & $ \pm$ 0.14 & $\pm$ 0.10 & $\pm$ 0.13\\
&&&&&&&&&& $\pm$ 0.13\\
$\Xi{_c(2645)^0}$ & $\Xi{_{c}^{+}} \pi^-$ & $1.82{_{-0.17}^{+0.09}}$ & $1.64{_{-0.16}^{+0.08}}$ \\
& $\Xi{_{c}^{0}} \pi^0$ & $0.90{_{-0.09}^{+0.04}}$  & $0.84{_{-0.08}^{+0.04}}$ \\
& Total & $2.72{_{-0.26}^{+0.13}}$ & $2.48{_{-0.24}^{+0.12}}$ & 2.9 & 2.8 & 2.53 & 3.12 & 1.83 & 3.03 & 2.35\\
&&&& $\pm$ 0.4 & $\pm$ 0.2 && $\pm$ 0.33 & $\pm$ 0.06 & $\pm$ 0.10 & $\pm$ 0.18\\
&&&&&&&&&& $\pm$ 0.13\\
$\Xi{_c(2790)^+}$ & $\Xi{_{c}^{\prime+}} \pi^0$ & $-$ & $2.28{_{-0.49}^{+0.55}}$ \\
& $\Xi{_{c}^{\prime0}} \pi^+$ & $-$ & $4.71{_{-1.04}^{+1.16}}$ \\
& Total & $-$ & $6.99{_{-1.53}^{+1.71}}$ & $8.0{_{-3.3}^{+4.7}}$ & $8.0{_{-3.3}^{+4.7}}$ & & & & & 8.9\\
&&&&&&&&&& $\pm$ 0.6\\
&&&&&&&&&& $\pm$ 0.8\\
$\Xi{_c(2790)^0}$ & $\Xi{_{c}^{\prime+}} \pi^-$ & $-$ & $4.79{_{-1.05}^{+1.18}}$ \\
& $\Xi{_{c}^{\prime0}} \pi^0$ & $-$ & $2.27{_{-0.51}^{+0.58}}$ \\
& Total & $-$ & $7.06{_{-1.56}^{+1.76}}$ & $8.5{_{-3.5}^{+5.0}}$ & $8.5{_{-3.5}^{+5.0}}$ &&&&& 10.0\\
&&&&&&&&&& $\pm$ 0.7\\
&&&&&&&&&& $\pm$ 0.8\\
$\Xi{_c(2815)^+}$ & $\Xi{_c(2645)^+} \pi^0$ & $1.58{_{-0.35}^{+0.39}}$ & $1.48{_{-0.32}^{+0.36}}$ \\
& $\Xi{_c(2645)^0} \pi^+$ & $3.11{_{-0.68}^{+0.77}}$ & $2.93{_{-0.64}^{+0.72}}$ \\
& Total & $4.69{_{-1.03}^{+1.16}}$ & $4.40{_{-0.97}^{+1.09}}$ & $3.4{_{-1.4}^{+2.0}}$ & $3.4{_{-1.4}^{+2.0}}$ & & 0.70 & 2.35 & & 2.43\\
&&&&&&& $\pm$ 0.04 & $\pm$ 0.93 && $\pm$ 0.20\\
&&&&&&&&&& $\pm$ 0.17\\
$\Xi{_c(2815)^0}$ & $\Xi{_c(2645)^+} \pi^0$ & $1.56{_{-0.34}^{+0.38}}$ & $1.57{_{-0.34}^{+0.39}}$ \\
& $\Xi{_c(2645)^0} \pi^+$ & $2.99{_{-0.68}^{+0.76}}$ & $3.07{_{-0.68}^{+0.76}}$ \\
& Total & $4.54{_{-1.00}^{+1.12}}$ & $4.64{_{-1.02}^{+1.15}}$ & $3.6{_{-1.5}^{+2.1}}$ & $3.6{_{-1.5}^{+2.1}}$ &&&&& 2.54\\
&&&&&&&&&& $\pm$ 0.18\\
&&&&&&&&&& $\pm$ 0.17\\
\noalign{\smallskip}\hline
\end{tabular}}
{\tiny{Ref. \cite{Cheng20151} $\rightarrow$ HHChPT(2015) PDG-2014 and\\
Ref. \cite{Cheng2007} $\rightarrow$ HHChPT(2007) PDG-2006.}}
\end{table}

\subsection{Strong Decays}

The study of heavy-light flavored baryons provides an excellent base to update the understanding of heavy quark symmetry of the heavy quark and the SU(3) chiral symmetry of the light quark. As we stated into the introduction the Heavy Hadron Chiral Perturbation Theory (HHChPT) describe a chiral Lagrangian in which both the symmetries are incorporated. Such a Lagrangian derived an expression of the partial decay rates (see Refs. \cite{Pirjol1997,Cheng2007,Cheng2015}) containing a strong couplings: $g_1$ and $g_2$ for the $P$-wave transitions and $h_2$ to $h_7$ for the $S$-wave transitions. And, the pion decay constant $f_{\pi}$ = 130.2 MeV is taken from PDG-2018 \cite{Tanabashi2018}.\\

\begin{table}
\caption{\label{tab:7}The ground state magnetic moments of the strange singly charmed baryons (in $ \mu_N $).}
\scalebox{0.9}{
\begin{tabular}{ccccccccccccccccccccccccccc}
\hline\noalign{\smallskip}
Baryon & Expression & Present & \cite{Wang2019} & \cite{Meng2018} & \cite{HCKim2019,Yang2018} & \cite{Shi2018} & \cite{Aliev2009} & \cite{Faessler2006} & \cite{Patel2008} & \cite{Sharma2010} & \cite{Can2015,Can2014} \\
\hline\noalign{\smallskip}
$\Xi{_{c}^{0}}$ & $\frac{2}{3}\mu_d + \frac{2}{3}\mu_s - \frac{1}{3}\mu_c$ & -1.011 & -0.96 & $-$ & $-$ & 0.192 & $-$ & 0.39 & -0.964 & 0.28 & $-$\\
&&&&&&(17)&&&&\\
$\Xi{_{c}^{+}}$ & $\frac{2}{3}\mu_u + \frac{2}{3}\mu_s - \frac{1}{3}\mu_c$ & 0.523 & $-$ & $-$ & $-$ & 0.235 & $-$ & 0.41 & 0.709 & 0.40 & $-$\\
&&&&&&(25)&&&&\\
$\Omega{_{c}^{0}}$ & $\frac{4}{3}\mu_s - \frac{1}{3}\mu_c$ & -0.842 & -0.96 & $-$ & 0.85 & -0.688 & $-$ & -0.85 & -0.958 & -0.90 & -0.688\\
&&&&& $\pm$ 0.05 &(31) &&&&& $\pm$ 0.031\\
$\Xi{_c(2645)^0}$ & $ \mu_d + \mu_s + \mu_c $ & -0.825 & -1.43 & -0.61 & -1.57 & $-$ & -0.68 & $-$ & -0.688 & -1.43 & $-$ \\
&&&&& $\pm$ 0.06 & & $\pm$ 0.18 &&&&\\
$\Xi{_c(2645)^+}$ & $ \mu_u + \mu_s + \mu_c $ & 1.319 & 0.72  & 1.28 & 0.90 & $-$ & 0.62 & $-$ & 1.513 & 1.59 & $-$\\
&&&&& $\pm$ 0.04 & & $\pm$ 0.68 \\
$\Omega{_c(2770)^0}$ & $ 2 \mu_s + \mu_c $ & -0.560 & -1.43 & -0.27 & -1.28 & $-$ & -1.68 & $-$ & -0.865 & -0.86 & -0.730 \\
&&&&& $\pm$ 0.08 & & $\pm$ 0.42 &&&& $\pm$ 0.023 \\
\hline\noalign{\smallskip}
\end{tabular}}
\end{table}

The strong decays of $\Xi{_c(2645)^+}$ and $\Xi{_c(2645)^0}$ are \textbf{governed} by the coupling $g_2$. In our previous work \cite{Gandhi2019}, we extract $g_2$ from the decays: $\Sigma_c(2455)^{++} \rightarrow \Lambda{_{c}^+} \pi^+$, $\Sigma_c(2455)^0 \rightarrow \Lambda{_{c}^+} \pi^-$, $\Sigma_c(2520)^{++} \rightarrow \Lambda{_{c}^+} \pi^+$ and $\Sigma_c(2520)^0 \rightarrow \Lambda{_{c}^+} \pi^-$ as 
\begin{equation}
\label{eqn:31} 
\mid g_2 \mid = 0.550{_{-0.027}^{+0.013}}, \hspace{0.4cm} 0.544{_{-0.018}^{+ 0.010}},\\
 \hspace{0.5cm} 0.561{_{-0.007}^{+ 0.005}}, \hspace{0.4cm} 0.570{_{-0.009}^{+ 0.007}}
\end{equation}

\noindent respectively. Here, we used their masses and respective decay widths from PDG-2018 \cite{Tanabashi2018}. Which are close to $\mid g_2 \mid_{2015}$ = $0.565{_{-0.024}^{+0.011}}$ and $\mid g_2 \mid_{2006}$ = $0.605{_{-0.043}^{+0.039}}$ and in accordance with the recent prediction $0.688{_{-0.035}^{+0.013}}$ of Ref. \cite{Kawakami2018}. The coupling $h_2$ controls the decay of first orbital excitation of isodoublet $\Xi{_c}$ baryon. The CDF Collaboration \cite{Aaltonen2011} studied $\Lambda_c(2595)^+$ resonance into $\Lambda{_{c}^+ \pi^+ \pi^-}$ final state and measured the strong coupling $h_2$ = 0.60 $\pm$ 0.07.\\

Expression of the strong decay rates of an isodoublet of $\Xi{_c(2645)}$, $\Xi{_c(2790)}$ and $\Xi{_c(2815)}$ are taken from the Refs. \cite{Pirjol1997,Cheng2007,Cheng2015}. The masses of these states obtained in the present study and from the PDG-2018 \cite{Tanabashi2018} are used to calculate their strong decay rates separately. They are listed into the third and the fourth column of the Table \ref{tab:6}. Here, in both the cases of masses we employed the same couplings $\mid g_2 \mid$ = $0.550{_{-0.027}^{+0.013}}$ and $h_2$ = 0.60 $\pm$ 0.07. Our results HHChPT(2019) with PDG-2018\cite{Tanabashi2018} update the strong decay rates of these baryonic states. Our calculated decay widths of isodoublet $\Xi{_c(2645)}$ are in agreement with HHChPT(2015) \cite{Cheng20151}, HHChPT(2007) \cite{Cheng2007} and the recent observations of Belle Collaboration \cite{Yelton2016}. The decay widths of an isodoublet states of $\Xi{_c(2790)}$ are smaller and for an isodoublet $\Xi{_c(2815)}$ our results are larger with respect to Refs. \cite{Cheng20151,Cheng2007,Yelton2016}. Such a differences are because of the selection of couplings the Ref. \cite{Cheng2007} used $\mid g_2 \mid$ = $0.605{_{-0.043}^{+0.039}}$ and $h_2$ = $0.437{_{-0.102}^{+0.114}}$ and the Ref. \cite{Cheng20151} used $\mid g_2 \mid$ = $0.565{_{-0.024}^{+0.011}}$ and $h_2$ = 0.63 $\pm$ 0.07.

\begin{table}
\caption{\label{tab:8}The radiative decay widths ($ \Gamma_\gamma $) of strange singly charmed baryons (in keV).}
\scalebox{1.0}{
\begin{tabular}{ccccccccccccccccccccccc}
\hline\noalign{\smallskip}
Decay & Present & \cite{Wang2019} & \cite{Jiang2015} & \cite{Wang20171} & \cite{Majethiya2009} & \cite{Bernotas2013} & \cite{Aliev2012} & \cite{Aliev2009} \\
\hline\noalign{\smallskip}
$\Xi_c(2645)^0 \rightarrow \Xi{_{c}^{0}} + \gamma $ & 0.776 & 1.84 & 0.36 & $-$ & 0.30 & 0.908 & 1.318 & 0.66 $\pm$ 0.41 & \\
$\Xi_c(2645)^+ \rightarrow \Xi{_{c}^{+}} + \gamma $ & 90.28 & 21.6 & 502 & $-$ & 63.32 & 44.3 & 152.4 & 52 $\pm$ 32 & \\
$\Omega_c(2770)^0 \rightarrow \Omega{_{c}^{0}} + \gamma$ & 3.726 & 0.32 & 4.82 & 0.89 & 2.02 & 1.07 & 1.439 & $-$ & \\
\hline\noalign{\smallskip}
\end{tabular}}
\end{table}

\subsection{Magnetic Moments}

To improve our understanding of the inner structure and geometrical shape of the hadrons the study of electromagnetic properties is an essential key. The magnetic moment of the hadron is precisely a function of its structure and of its parameters such as spin, flavor, charge and the effective masses of the bound quarks. It's expression can be written in the form of expectation value as \cite{Gandhi2018,Majethiya2009}, 

\begin{equation}
\label{eqn:32} 
\mu_B = \sum_{q} \left\langle \Phi_{sf} \middle| \hat{\mu}_{qz} \middle| \Phi_{sf} \right\rangle;  \hspace{0.5cm} q=u,d,s,c.
\end{equation}

Here, the $\mu_B$ and the $\Phi_{sf}$ denotes the magnetic moment and the spin-flavor wave function of the participating baryon. The $\hat{\mu}_{q_{z}}$ is the $z$-component of the magnetic moment operator of the individual quark given by,

\begin{equation}
\label{eqn:33} 
\hat{\mu}_{qz} = \frac{e_q}{2 m{_{q}^{eff}}} \cdot \sigma_{qz};
\end{equation} 

\noindent where $ e_q $ is the charge and $\sigma_{q_{z}}$ is the $z$-component of the constituent quark spin. The $m{_{q}^{eff}}$ is the effective quark mass and is,

\begin{equation}
\label{eqn:34} 
m{_{q}^{eff}} = m{_{q}} \left({1 + \frac{\left\langle H \right\rangle}{\sum\limits_{q} m{_{q}}}}\right). 
\end{equation}

\noindent Eq. \ref{eqn:34} gives the mass of the bound quark inside the baryons by taking into account its binding interactions with other two quarks. The Hamiltonian $ {\left\langle  H \right\rangle}$ is obtained by taking the mass difference of the measured or predicted baryon mass $M$ with the sum of the masses of the three constituent quarks $\sum\limits_{q} m{_{q}}$ of the participating baryon, i.e., $ {\left\langle  H \right\rangle} = M - {\sum\limits_{q} m{_{q}}} $.\\

For example: in order to determine the magnetic moment of $\Omega{_{c}^0}$ baryon we write the Eq. \ref{eqn:32} as,

\begin{equation}
\label{eqn:35} 
\mu_{\Omega{_{c}^0}} = \left \langle\Phi_{sf_{\Omega{_{c}^0}}} \right \vert \hat{\mu}_{qz} \left\vert \Phi_{sf_{\Omega{_{c}^0}}} \right\rangle
\end{equation}

\noindent here the spin-flavor wave function ($\Phi_{sf}$) of $\Omega{_{c}^0}$ baryon is obtained by taking a combination of flavor and spin part wave function as,

\begin{equation}
\label{eqn:36} 
\left\vert \Phi_{sf_{\Omega{_{c}^0}}} \right\rangle  = \frac{1}{\sqrt{6}} \left(s\uparrow s\downarrow c\uparrow + s\downarrow s\uparrow c\uparrow - 2 s\uparrow s\uparrow c\uparrow \right)
\end{equation}

\noindent Therefore, the Eq. \ref{eqn:35} will be 

\begin{equation}
\label{eqn:37}
\mu_{\Omega{_{c}^0}} = \frac{1}{6} \left \langle s\uparrow s\downarrow c\uparrow + s\downarrow s\uparrow c\uparrow - 2 s\uparrow s\uparrow c\uparrow \right \vert \hat{\mu}_{qz} \left\vert s\uparrow s\downarrow c\uparrow + s\downarrow s\uparrow c\uparrow - 2 s\uparrow s\uparrow c\uparrow \right\rangle.
\end{equation}

\noindent Following the orthogonality of quark flavor and spin states, for example $ \left\langle s\uparrow s\downarrow c\uparrow \vert s\uparrow s\uparrow c\downarrow \right\rangle = 0 $, the expression for the $\Omega{_{c}^0}$ magnetic moment reduces to  

\begin{equation}
\label{eqn:38}
\mu_{\Omega{_{c}^0}} = \frac{1}{6} \left \langle s\uparrow s\downarrow c\uparrow \right \vert \hat{\mu}_{qz} \left\vert s\uparrow s\downarrow c\uparrow \right\rangle \\
 + \frac{1}{6} \left \langle s\downarrow s\uparrow c\uparrow \right \vert \hat{\mu}_{qz} \left\vert s\downarrow s\uparrow c\uparrow \right\rangle \\
 + \frac{4}{6} \left \langle s\uparrow s\uparrow c\downarrow \right \vert \hat{\mu}_{qz} \left\vert s\uparrow s\uparrow c\downarrow \right\rangle.
\end{equation}

Expectation values of the $z$-component of the magnetic moment operator of the strange and charm quarks are,  

\begin{equation}
\left \langle s\uparrow \right \vert \hat{\mu}_{sz} \left\vert s\uparrow \right\rangle = +\mu_s, \hspace{0.2cm} \left \langle s\downarrow \right \vert \hat{\mu}_{sz} \left\vert s\downarrow \right\rangle = -\mu_s, \hspace{0.2cm}
\left \langle c\uparrow \right \vert \hat{\mu}_{cz} \left\vert c\uparrow \right\rangle = +\mu_c, \hspace{0.2cm} \left \langle c\downarrow \right \vert \hat{\mu}_{cz} \left\vert c\downarrow \right\rangle = -\mu_c.\\     
\end{equation}

\noindent Therefore, Eq. \ref{eqn:38} becomes 

\begin{equation}
\label{eqn:39}
\mu_{\Omega{_{c}^0}} = \frac{1}{6} \left(\mu_s - \mu_s + \mu_c \right) + \frac{1}{6} \left(- \mu_s + \mu_s + \mu_c \right) + \frac{4}{6} \left( \mu_s + \mu_s - \mu_c \right),
\end{equation}

\noindent and this gives the quark model prediction for the ground state magnetic moment of the $\Omega{_{c}^0}$ baryon as,   

\begin{equation}
\label{eqn:40} 
\mu_{\Omega{_{c}^0}} = \frac{4}{3} \mu_s - \frac{1}{3} \mu_c.
\end{equation}

In the same way, we also determine an expression of the magnetic moment of the other strange singly charmed baryons, as reported in Table \ref{tab:7}. Using the constituent quark masses from section \ref{sec:2} and the masses of the strange singly charmed baryons from Table \ref{tab:2} for $J^P = {\frac{1}{2}}^+$, we calculate their magnetic moment and compare them with other theoretical predictions, as reported in Table \ref{tab:7} .

\subsection{Radiative Decays}

Electromagnetic transition strength of the singly charmed baryons is weaker and has not a phase space restriction than that of the pion transitions in the strong decay \cite{Cheng20151}. But some radiative decay modes contribute significantly to the total branching fractions of the singly charmed baryon. The radiative decay width can be expressed in terms of the radiative transition magnetic moments $(\mu_{B_{C}^{\prime}\rightarrow B_{c}})$ in $\mu_N$ and photon energy ($k$) as discussed in \cite{Gandhi2018,Majethiya2009},  

\begin{equation}
\label{eqn:41} 
\Gamma_{\gamma} = \frac{k^3}{4\pi}\frac{2}{2J+1}\frac{e}{m{_{p}^2}}\mu{_{{B{_{c}}} \rightarrow {B{_{c}^{\prime}}}}^2}
\end{equation}

\noindent where $m_p$ is the proton mass and $J$ represents the total angular momentum of the initial baryon $B_c$.\\

The transition magnetic moment is determined by following the same procedure as we discussed in the above subsection. For the transition $ \Omega_c(2770)^0 \rightarrow \Omega{_{c}^{0}} $; we write

\begin{equation}
\label{eqn:42} 
\mu_{\Omega_c(2770)^0 \rightarrow \Omega{_{c}^{0}}} = \left \langle\Phi_{sf_{\Omega_c(2770)^0}} \right \vert \hat{\mu}_{qz} \left\vert \Phi_{sf_{\Omega{_{c}^{0}}}} \right\rangle
\end{equation}

\noindent the spin-flavour wave functions $(\Phi_{sf})$ of $\Omega_{c}(2770)^0$ and $ \Omega{_{c}^{0}} $ states are,

\begin{equation}
\label{eqn:43} 
\left\vert \Phi_{sf_{\Omega_c(2770)^0}} \right\rangle  = \left(ssc\right) \cdot \left(\frac{1}{\sqrt{3}}(\uparrow\uparrow\downarrow+\uparrow\downarrow\uparrow+\downarrow\uparrow\uparrow)\right)
\end{equation}

\begin{equation}
\label{eqn:44} 
\left\vert \Phi_{sf_{\Omega{_{c}^{0}}}} \right\rangle  = \left(ssc\right) \cdot \left(\frac{1}{\sqrt{6}}(2\uparrow\uparrow\downarrow-\uparrow\downarrow\uparrow-\downarrow\uparrow\uparrow)\right).
\end{equation}

Now repeating the procedure as we have done in the above subsection, this gives the transition magnetic moment (in $\mu_N$) as, 

\begin{itemize}
\item $\mu_{\Omega_c(2770)^0 \rightarrow \Omega{_{c}^{0}}}$ $~~~~~$ $2\sqrt{\frac{2}{3}} \left (\mu_s - \mu_c \right)$ $~~~~~$ -1.499.
\end{itemize}

In the same way, we determine the transition magnetic moment of the other strange singly charmed baryonic states,

\begin{itemize}
	\item $\mu_{ \Xi_c(2645)^0 \rightarrow \Xi{_{c}^{0}}}$ $~~~~~$ $\sqrt{\frac{2}{3}} \left (\mu_d - \mu_s \right)$ $~~~~~$ -0.176
	\item $\mu_{ \Xi_c(2645)^+ \rightarrow \Xi{_{c}^{+}}}$ $~~~~~$ $\sqrt{\frac{2}{3}} \left (\mu_u - \mu_s \right)$ $~~~~~$ 1.864. 
\end{itemize}

Using these radiative transition magnetic moments, the radiative decay widths are calculated (in keV). Our results are listed in Table \ref{tab:8} and compared with other theoretical predictions.

\section{Summary}
\label{sec:5}

In the present study, the excited state masses of the strange singly charmed baryons are calculated in the non-relativistic framework of hypercentral Constituent Quark Model (hCQM). A screened potential is used as a confining potential with the first order correction to see the relativistic effect in the heavy-light baryonic systems. Our results are listed in Tables \ref{tab:2}-\ref{tab:5} with other theoretical predictions and the available experimental measurements. Moreover, our results are plotted on Regge line according to their quantum number in both $(n_r, M^2)$  and $(J, M^2)$ planes as shown in Figs. \ref{fig:1}-\ref{fig:6}. That help to assign the possible  $J^P$ value of the experimentally observed unknown states. Here, the spectroscopy and the Regge trajectories studies identified the states such as:\\

\noindent $\bullet$ $\Xi{_c}$ baryon\\

an isodoublet of $\Xi{_c(2970)}$ and $\Xi{_c(3080)}$ states are assigned as a $2S$ state with $J^P$ = ${\frac{1}{2}}^+$ and ${\frac{3}{2}}^+$ respectively; and the $\Xi{_c(3123)^+}$ state is predicted as a $1D$-state with $J^P$ = ${\frac{3}{2}}^+$ for a spin $S$ = ${\frac{1}{2}}$.\\

\noindent $\bullet$ $\Omega{_c^0}$ baryon\\

the $\Omega{_c}(3000)^0$ state identified as a $1P$-state with $J^P$ = ${\frac{1}{2}}^-$ and the $\Omega{_c}(3119)^0$ state may be $2S$-state with $J^P$ = ${\frac{1}{2}}^+$.\\

The experimentally observed strange baryons resonances $\Xi{_c}(2930)$, $\Xi{_c(3055)^+}$, $\Omega{_c}(3050)^0$, $\Omega{_c}(3066)^0$ and $\Omega{_c}(3090)^0$ are not identified with their spin-parity in this work.\\

\noindent The strong decay rates of the isodoublet baryonic states $\Xi{_c(2645)}$, $\Xi{_c(2790)}$ and $\Xi{_c(2815)}$ are analyzed in the framework of Heavy Hadron Chiral Perturbation Theory (HHChPT). The decays of the isodoublet $\Xi{_c(2645)}$ is driven by the strong coupling $g_2$ and $h_2$ controls the decays of $\Xi{_c(2790)}$ and $\Xi{_c(2815)}$ isodoublet states. Here, we used $\mid g_2 \mid$  = $0.550{_{-0.027}^{+0.013}}$ extract in our previous work \cite{Gandhi2019} by using the masses from PDG-2018 \cite{Tanabashi2018} and $h_2$ = 0.60 $\pm$ 0.07 was taken from the CDF Collaboration \cite{Aaltonen2011}. Hence, using these couplings and the baryons masses from PDG-2018 \cite{Tanabashi2018} we update the values of their strong decay rates. Moreover, the magnetic moments and the transition magnetic moments of the ground state of strange singly charmed baryons are calculated in the constituent quark model. The radiative decay rates based on the transition magnetic moments are determined. Our results of magnetic moments and the radiative decay rates are compatible with other theoretical predictions.\\

 The obtained mass spectra of strange singly charmed baryons are used for the study of their Regge trajectories, strong decays, magnetic moments, transition magnetic moments and radiative decay widths. This study will help experimentalists as well as theoreticians to understand the dynamics of strange singly charmed baryons. So we foresee to extend this scheme for the calculations of the strange and non strange singly bottom baryons.

\end{document}